\documentclass[preprint,12pt]{elsarticle}

\usepackage{graphicx}
\usepackage{amssymb}

\usepackage{lineno}
\usepackage{textcomp}
\usepackage{gensymb}

\newcommand{\comment}[1]{}

\usepackage{xcolor}
\usepackage{graphicx}
\usepackage{float}

\usepackage{subcaption}
\usepackage{mhchem}
\newcommand\isotope[2]{\textsuperscript{#2}#1}
\graphicspath{{./images/}}

\journal{NIM Section A}

\begin{document}

\begin{frontmatter}


\title{A novel technique for real-time ion identification and energy measurement for in situ space instrumentation}



\author[KS]{F. Gautier}
\author[NASA]{A. Greeley}
\author[NASA]{S. G. Kanekal}
\author[KS]{T. Isidori}
\author[KS]{G. Legras} 
\author[KS]{\\ N. Minafra}
\author[KS]{A. Novikov\corref{cor1}}
\cortext[cor1]{Corresponding author.Tel.: +1 785 864 4626; Fax: +1 785 864 5262}
\ead{alexander.novikov@ku.edu}
\author[KS]{C. Royon}
\author[Q]{Q. Schiller}

\address[KS]{Department of Physics and Astronomy, University of Kansas, Lawrence, KS 66045, USA}
\address[NASA]{Heliophysics Division, NASA Goddard Space Flight Center, Greenbelt, MD 20771, USA }
\address[Q]{Space Science Institute, Boulder, CO 80301, USA }


\begin{abstract}

The AGILE (Advanced enerGetic Ion eLectron tElescope) project focuses on the development of a compact low-cost space-based instrument to measure the intensities of charged particles and ions in space.
Using multiple layers of fast silicon sensors and custom front-end electronics, the instrument is designed for real-time particle identification of a large variety of elements from H to Fe and spanning energies from 1 to 100 MeV per nucleon.
The robust method proposed in this work uses key defining features of electronic signals generated by charged particles (ions) traveling through silicon layers to reliably identify and characterize particles in situ.
AGILE will use this real-time pulse shape discrimination technique for the first time in space based instrumentation.

\end{abstract}


\end{frontmatter}



\section{Introduction}

\label{S:intro}
The technical challenge presented by the identification of detected particles (or ions) and the precise measurement of their energy is one of the core aspects of many physics research areas. 
Nuclear, particle physics, and applied medical studies (e.g. measurement of the equivalent dose received by the human body) often share the need for designing robust methods to discriminate amongst particle species. 
Space physics experiments also strongly rely on these techniques for investigating intensities and energy spectra of charged particles (particularly ions) originating from solar, distant heliospheric, and galactic sources.
In situ particle identification is necessary to understand ion  production, energy distribution, and abundance in the universe as well as the underlying physical processes that energize and transport them throughout space. 
Near the Earth and in the heliosphere, dynamics of radiation belt electrons, solar energetic particles and cosmic ray energization and transport are prime examples of charged particle processes that remain to be fully characterized.
 Real-time discrimination of incoming particles and extraction of their intrinsic features can lead to improvements of existing theoretical models. Furthermore, understanding the effects of charged particles on the human body during space activities and future exploration (e.g. Mars) is a critical aspect of Space Weather.  \par

The main goal of the NASA-funded project AGILE (Advanced enerGetic Ion eLectron tElescope) is the development of a compact, low mass, low power, and low cost instrument capable of detecting and identifying ions from H to Fe over a wide energy range, spanning 1-100 MeV per nucleon (MeV/n). 
AGILE includes full read-out and front-end electronics, which allows on-board, real-time discrimination of incident particles. Direct communication between the front-end and the detection layers prevents sending vast amounts of raw data back to Earth, which makes AGILE a prime candidate for small missions, like CubeSats, which are miniaturized satellites comprised of cube units (termed U) with dimensions of $10 cm \times 10 cm \times 10 cm$   \cite{heidt2000cubesat, helvajian2008small, poghosyan2017cubesat} . Several CubeSat missions have been launched and  have made significant contributions to the space science. For example Li et al. \cite{li2017measurement} have demonstrated that a significant source of energetic electrons in the inner radiation belt results from the so-called CRAND, i.e., cosmic ray albedo neutron decay. Their results were based on measurements from the REPTile instrument on the CSSWE CubeSat, which also provided complementary measurements for a major NASA mission, the Van Allen Probes, and helped establish a hitherto unknown feature of the Earth’s outer radiation belts, namely the “impenetrable barrier” \cite{baker2014impenetrable}. Crew et al. have characterized electron microburst precipitation from the radiation belts using measurements by sensors onboard the FIREbird CubeSat  \cite{crew2016first}. \par
This work presents a new approach to identify ion species (ID) based on a Pulse Shape Discrimination (PSD) technique \cite{PSD, Carboni} which will be used for the first time in space based instrumentation. The key difference and improvement of the PSD method over the "classic" $\Delta E-E$ technique \cite{goulding1975identification} is that the former utilizes all the information contained in the pulse shape, i.e, both amplitude (pulse height) and its temporal evolution. Furthermore, the PSD method can determine energy and ID using the pulse signal from a single detector. An additional advantage of the PSD method is its capability to identify ions of a lower energy threshold (e.g., 6 MeV/n as opposed to 20 MeV/n for calcium ions as demonstrated by Carboni et al. \cite{Carboni}). While the $\Delta E-E$ method has been successfully used for space based experiments for decades \cite[e.g.][]{stone1977, stone1998cosmic}, it relies solely on pulse height and requires a stack of detectors. Recent advances in electronics, i.e., wave form samplers such as the PSEC4 used in AGILE (Section \ref{subsection:psec}) have enabled the use of PSD technique in space-based instrumentation. It also benefits from very fast silicon detectors that produce signals with a duration of a few nanoseconds. \par
 Following this introduction, the second section describes the discrimination method for ion identification and energy measurement. The third section describes the instrument architecture and the PSD method validation using detailed and complete simulations.
Section 4 shows the simulated performance of the proposed method, followed by the conclusion.

\section{Charged Particles Identification Based on Pulse Shape Discrimination}

\subsection{Objectives}

The AGILE project currently focuses on the robust identification of H to Fe in an energy range of 1-100 MeV/n with an energy resolution of $\sim10\%$, however this method may be extended for broader particle and energy ranges. 
Due to the limitation of the amount of data that can be downlinked from space, the discrimination method needs to be performed in real time onboard a satellite. Furthermore, since CubeSats have size and power restrictions, the detection system must be compact and consume relatively low power.
The algorithms used to identify an ion and measure its energy must be simple enough to be implemented in on-board electronics (e.g ASIC or FPGA), thereby drastically reducing the amount of data that needs to be sent to the ground stations.

\subsection{General Approach}
For space and ground based experiments, one of the most widely used discrimination techniques is $\Delta E-E$ which uses the fact that energy  deposition in at least two consecutively stacked detector elements (the first one is usually much thinner than the second) is different for various particle species and requires a particle to pass thorough the first element and stop in the second one. However, the use of multiple detectors makes the detection system more complex and increases the instrument noise, particularly for thin detectors. The PSD technique on the other hand, can be applied even to a single detector, and is a very promising solution for an instrument like AGILE. The method relies on the fact that different particles deposit different amounts of energy along their tracks through the detector medium (Bragg curve, $\frac{dE}{dx}(x)$ - energy loss $E$ per unit path length $x$). This results in various signal shapes (in both amplitude and time domains) that can be used for particle discrimination and energy measurements. In particular, it is known that a single solid state detector (e.g. Si) can be used for PSD if a particle loses all of its energy within the detector medium \cite{mutterer2000}. \par
However, in order to fully exploit the PSD method, very fast detection systems are required (detectors, front-end electronics, and samplers), since the typical signal duration in a Si-detector with a thickness of a few hundreds of microns is of the order of 10-100 ns \cite{leroy1999study}. \par

\begin{figure}[H]
    \centering
    \includegraphics[width=1.0\textwidth]{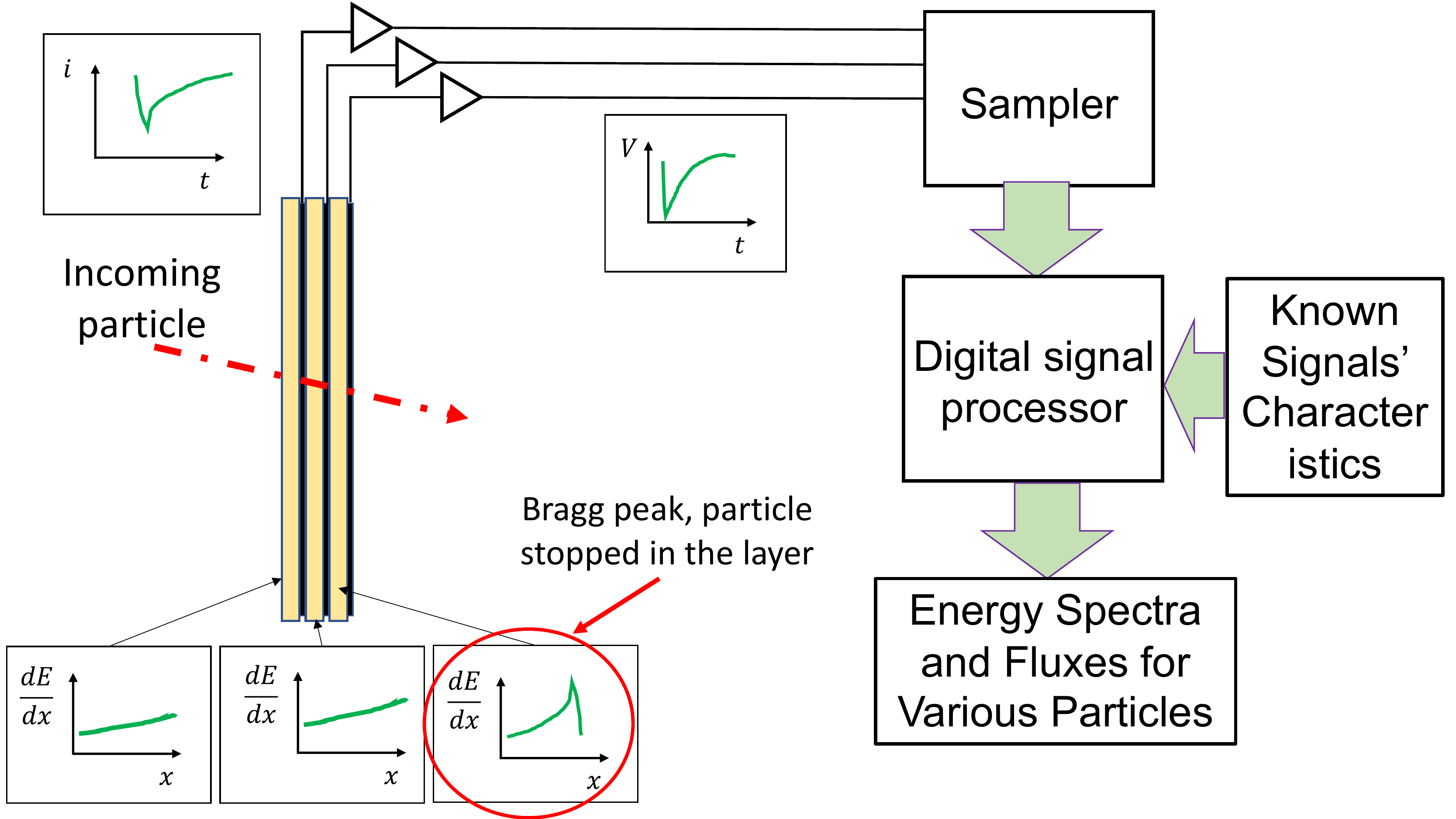}
    \caption{Simplified schematic of PSD method in a 3 layers configuration: a particle stops in the third layer (Bragg peak) while passing through the first and second layers. The signals from each layer are processed by fast analog read-out electronics and fed into a fast high-resolution sampler. 
    The sampler digitizes the entire signal corresponding to the energy deposition profile in a specific layer. The sampler is followed by a digital signal processor which compares the characteristics of incoming signals with a database of characteristics corresponding to specific kinds and energies of particles. If there is a match, the particle can be identified and its energy can be measured.}
    \label{fig:schematic}
\end{figure}

In the current study, a stack (3 layers, 300 $\mu$m each) of fast $p$-type silicon detectors (Section \ref{subsection:silicon_detectors}), fast analog electronics (Section \ref{subsection:amps}), and digitizing and read-out ASICs (Section \ref{subsection:psec}) are used. This configuration allows the electronics to reconstruct the different signals produced by the energy deposited in all detector layers with high precision. Careful analysis of the signal from the layer where a particle stops (Bragg peak) yields a set of simple characteristics, viz. rise time and amplitude, which are unique for a specific ion (atomic number Z and mass number M) and its energy, as will be shown in the following sections. Thus, the energy and species of the particle can be determined simultaneously. 
A simplified schematic of the approach is shown in Fig. \ref{fig:schematic}.

\subsection{Pulse Shape Discrimination Algorithm}
\label{subsection:PSDAlgorithm}
\begin{figure}[!htp] 
    \centering
    \includegraphics[width=1.25\textwidth]{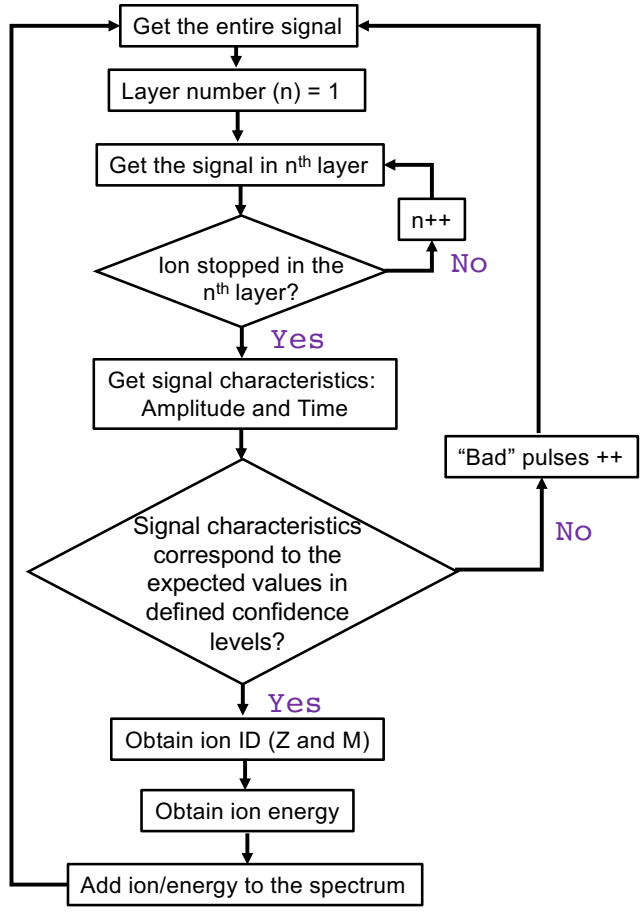}
    \caption{A simplified flowchart of the PSD algorithm implemented in a FPGA/ASIC in the case of a multi-layer silicon detector}
    \label{fig:flowchart_simplified}
\end{figure}

As mentioned above, the discrimination method will be implemented in a resource constrained environment (i.e. on satellites) and proceeds as follows (a simplified flowchart is shown in Fig. \ref{fig:flowchart_simplified}):

\begin{itemize}
    \item Obtain and digitize the signals from all 3 detector layers with high time and amplitude resolution (Section \ref{subsection:psec});
    \item Identify the layer where the particle stopped by using a specific trigger (Section \ref{subsection:range});
    \item Evaluate the key signal characteristics (namely the amplitude and the time) in the stopping layer;
    \item Evaluate the particle ID (Z and M) by using the unique dependence of the pulse amplitude and the rise time (Section \ref{ss:signalCharacteristics} and Fig. \ref{fig:IDdiscrimination});
    \item Estimate the energy of this particle knowing the particle ID by using the signal amplitude dependence on the energy (see Fig. \ref{fig:Ediscrimination});
    \item Consistency check\footnote{The signals that cannot be used to determine particle ID and energy (there is no match between the calculated signal characteristics and the expected values within a predefined confidence level) are considered to be ``bad" pulses. They contribute to the detector dead time and inefficiencies in the particle fluxes.}. 
\end{itemize}

 Based on the cumulative information about the detected particle IDs and energies, energy spectra and differential fluxes can be obtained for each specific ion. \par

\subsection{Simulation}
\label{subsection:simShort}
In order to ascertain the expected performance of the PSD method, detailed simulations with a configuration of three layers of silicon detectors were performed. The main steps of the simulations are the following:
\begin{enumerate}
\item Simulate the energy deposition profiles in each layer of the Si detector using GEANT4 software \cite{geant4};
\item Simulate the detector response (output signal: I(t)) by passing Geant4 energy deposition profiles to Weightfield2 software \cite{WF2}, a simulation tool for silicon and diamond detectors that was adapted to AGILE;   
\item Simulate the read out electronics (amplifier output: V(t)) by passing the Weightfield2 signal to the circuit simulator LTspice \citep{ltspice}.

\end{enumerate}

The simulation uses a random distribution of incident particles: ion type (H-Fe) and energy (1 MeV/n - 100 MeV/n) were randomly generated using a uniform distribution and then passed to the simulation chain described above. 

Full details of the simulation can be found in \ref{section:simulation}. It should be noted that these results can be extrapolated to an arbitrary number of silicon layers.

\subsection{Stopping Layer}
\label{subsection:range}

\begin{figure}[!htp]
    \hspace*{-8mm}
    \includegraphics[width=1.1\textwidth]{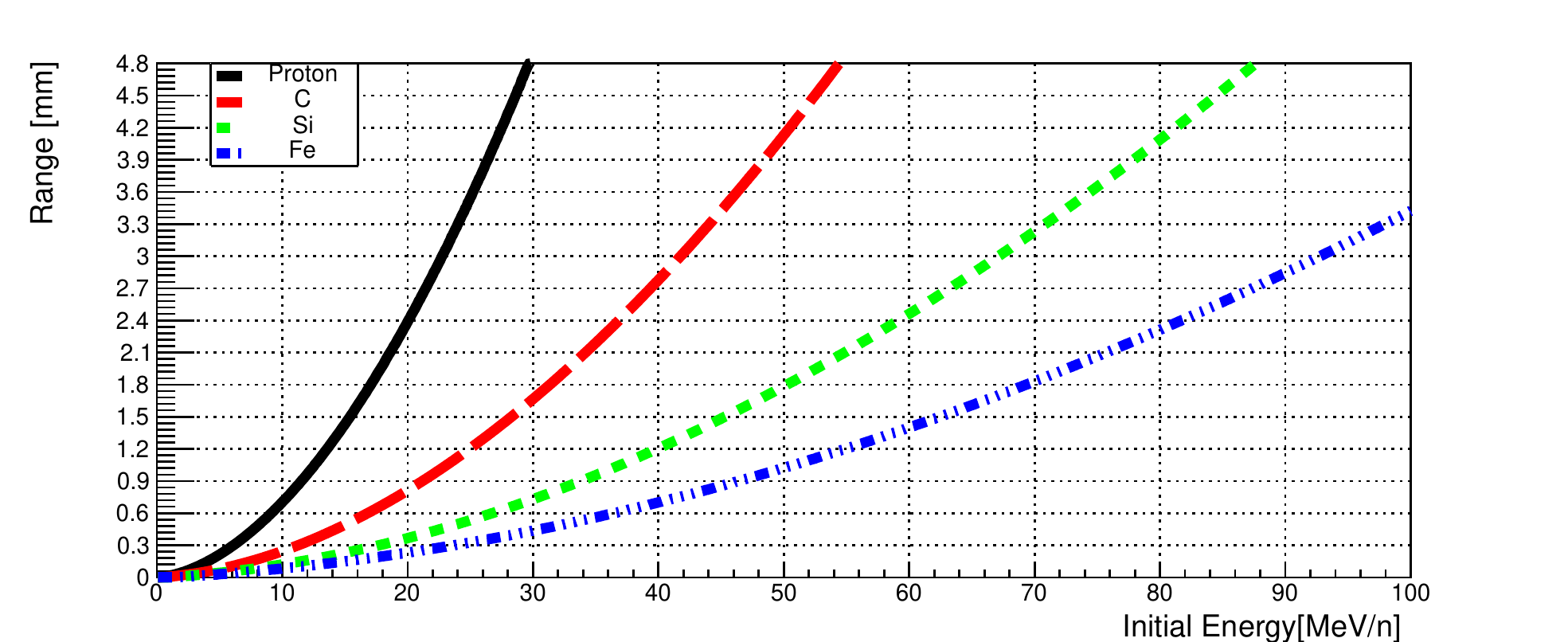}
    \hspace{4mm}
    \caption{Ranges in Si vs initial energy of ions for proton (black full line), C (red dashed line), Si (green dotted line) and Fe (blue dash-dotted line) are plotted for initial energy from 1 to 100 MeV/n}
    \label{fig:range}
\end{figure}

When a given incoming particle enters a multi-layer Si-detector stack, it is stopped in a given layer depending on its energy.
The energy and particle type is dependent on the thickness of each layer and the properties of the detector medium (Si). 
Fig. \ref{fig:range} shows the stopping range of particles depending on their initial incident energy.  (see \ref{subsection:silicon_detectors}).

As can be seen in Fig. \ref{fig:range}, the penetration depth of a particle is dependent on both the incident energy and its type.
For example, a heavy Fe ion will stop in the third layer of the detector when its energy is in the range of 36-46 MeV/n while for a proton this range is 9-11 MeV/n.

This method focuses on the particles with energies that are low enough to be stopped in one of the layers comprising the detector stack. Specific triggers will be used for each layer to determine if a given particle has deposited some energy in a given layer and thus determine the layer where it stopped. 
\subsection{Key Characteristics of the Signal}
\label{ss:signalCharacteristics}

Once the layer in which a particle stops is known, the key characteristics that allow particle identification and energy measurements can be defined.

\begin{figure}[!htp]
    \vspace{-3cm}
    \centering
    \begin{subfigure}{1\textwidth}
        \centering
        \includegraphics[width=1\textwidth]{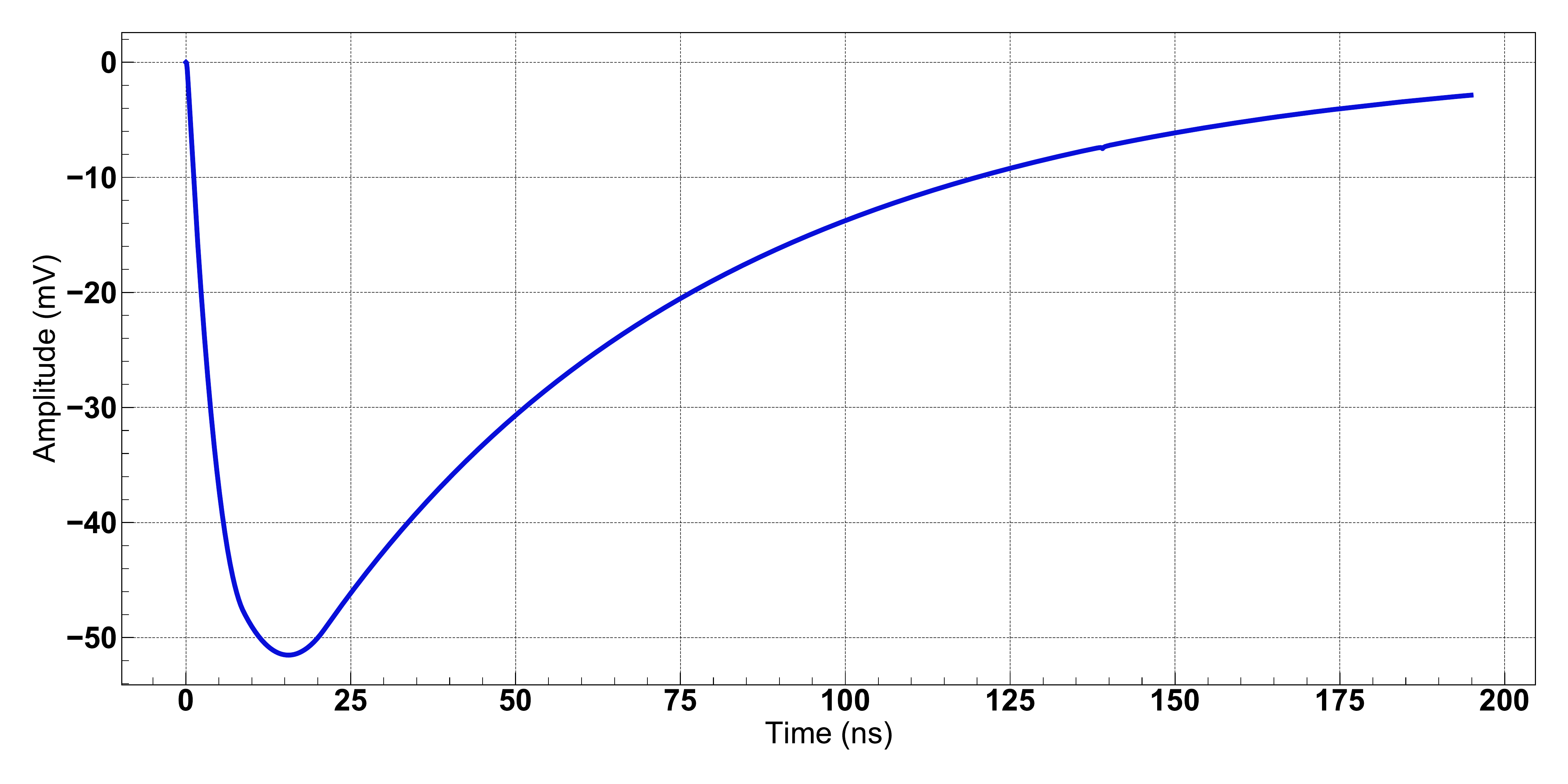}
        \caption{Layer 1.}
    \end{subfigure}
    \begin{subfigure}{1\textwidth}
        \centering
        \includegraphics[width=1\textwidth]{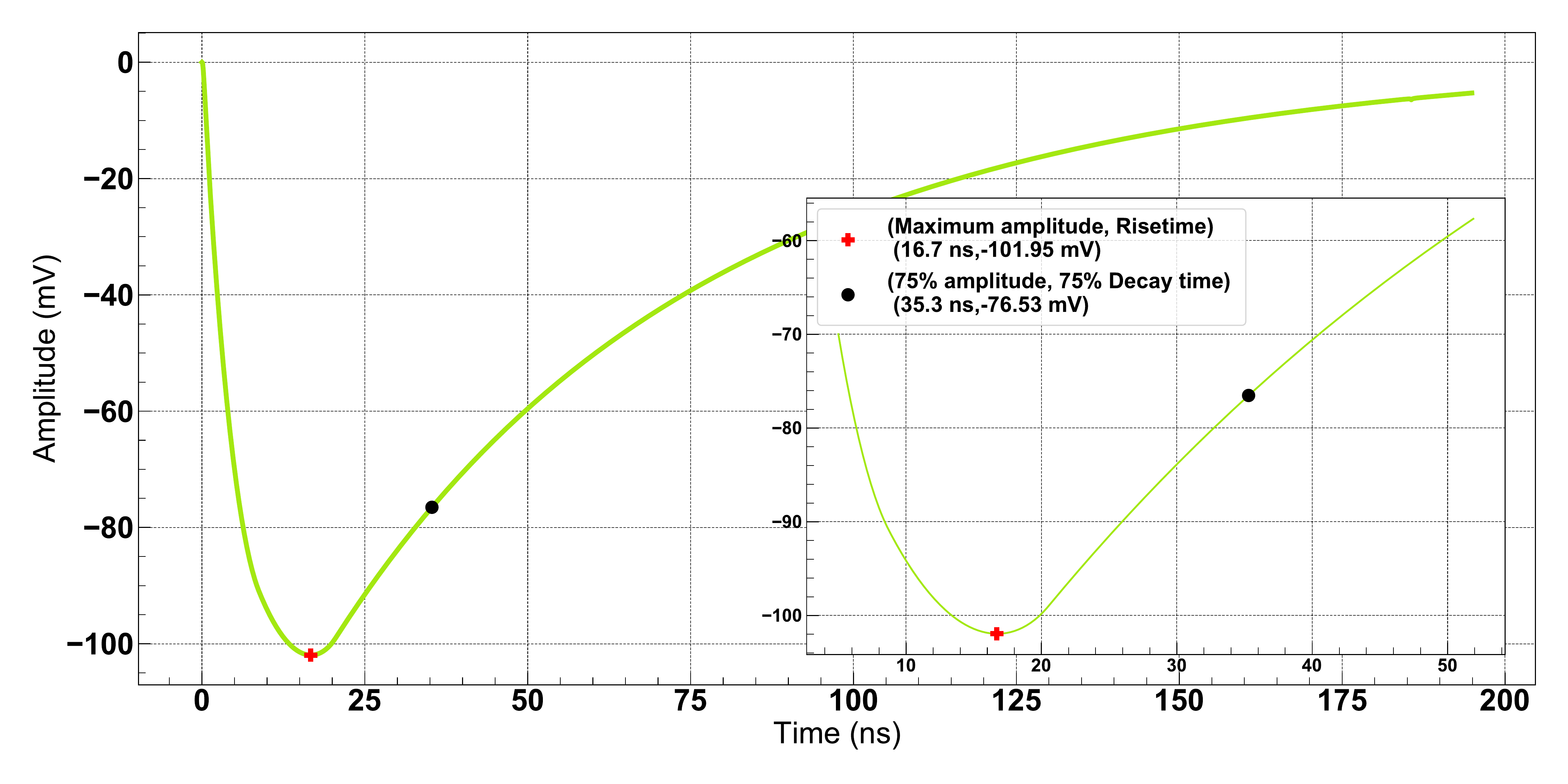}
        \caption{Layer 2.}
    \end{subfigure}
    \begin{subfigure}{1\textwidth}
        \centering
        \includegraphics[width=1\textwidth]{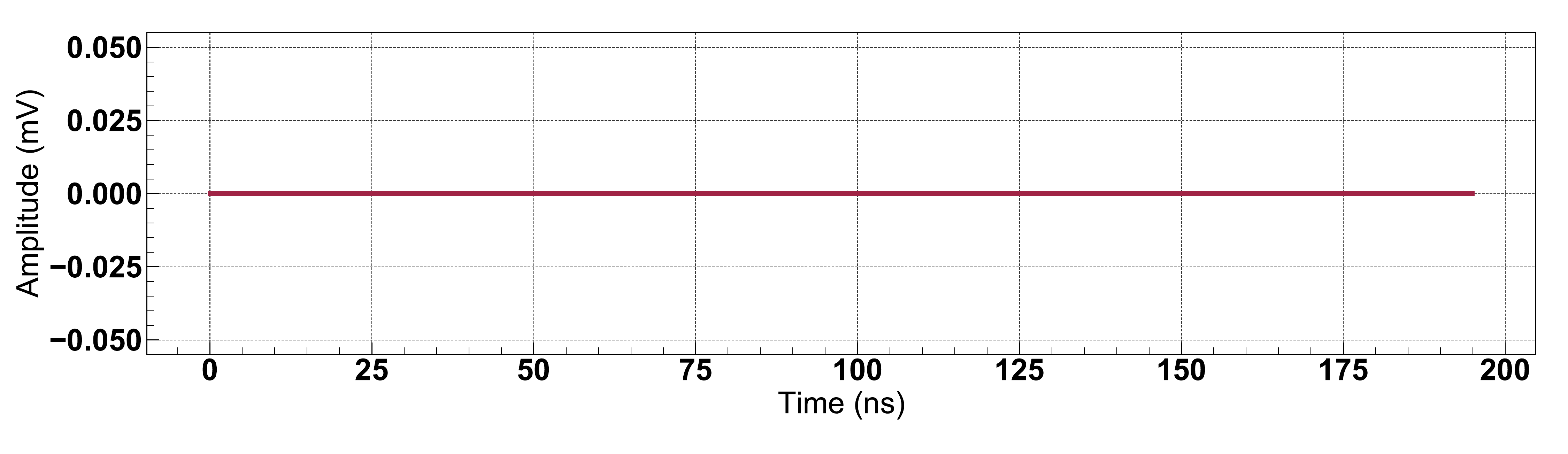}
        \caption{Layer 3.}
    \end{subfigure}
    \caption{Simulated signals in a stack of three 300 $\mu$m layers of Si produced by a 20 MeV/n oxygen ion that stops in the second layer. Each layer is respectively presented in Figures a), b) and c). The key characteristics of the pulse (Maximum Amplitude, Rise Time, and 75\% Decay Time) are indicated on the stopping layer (Fig b)) where an inset of the zoomed-in peak is included.}
    \label{fig:ex_lowgain}
\end{figure}

As an example, Fig. \ref{fig:ex_lowgain} shows the simulated signals from an event produced by an oxygen ion with an energy of 20 MeV/n in a stack of 3 layers. The particle deposits energy in the first layer then stops in the second since no energy is measured in the 3rd layer.
The key signal characteristics are the maximum amplitude (-101.95 mV in this example), the rise time (16.7 ns), and the 75\% decay time (time when the pulse amplitude decreases to 75\% of its maximum value, 35.3 ns). 

  In a fully depleted silicon detector where a particle stops completely, its ID (Z and M) and energy can be obtained by measuring the pulse rise time (proportional to the charge collection time) and its amplitude (proportional to the total charge collected) \cite{mutterer2000, england89, pausch94}. Our current studies are primarily based on these two characteristics, described in more detail below. 

\subsubsection*{Pulse Rise Time}
When a particle completely stops in a detector, the rise time is proportional to the total duration of the charge collection (both electron and hole components) and can be used as an indicator of the depth at which the particle stops. Since this depth is different for different particles (Z and M) at the same given energy, the rise time along with the pulse amplitude (proportional to the particle energy) can be used for robust particle identification \cite{england89}. 
\begin{figure}[!htp]
    \centering
    \includegraphics[width=0.85\textwidth]{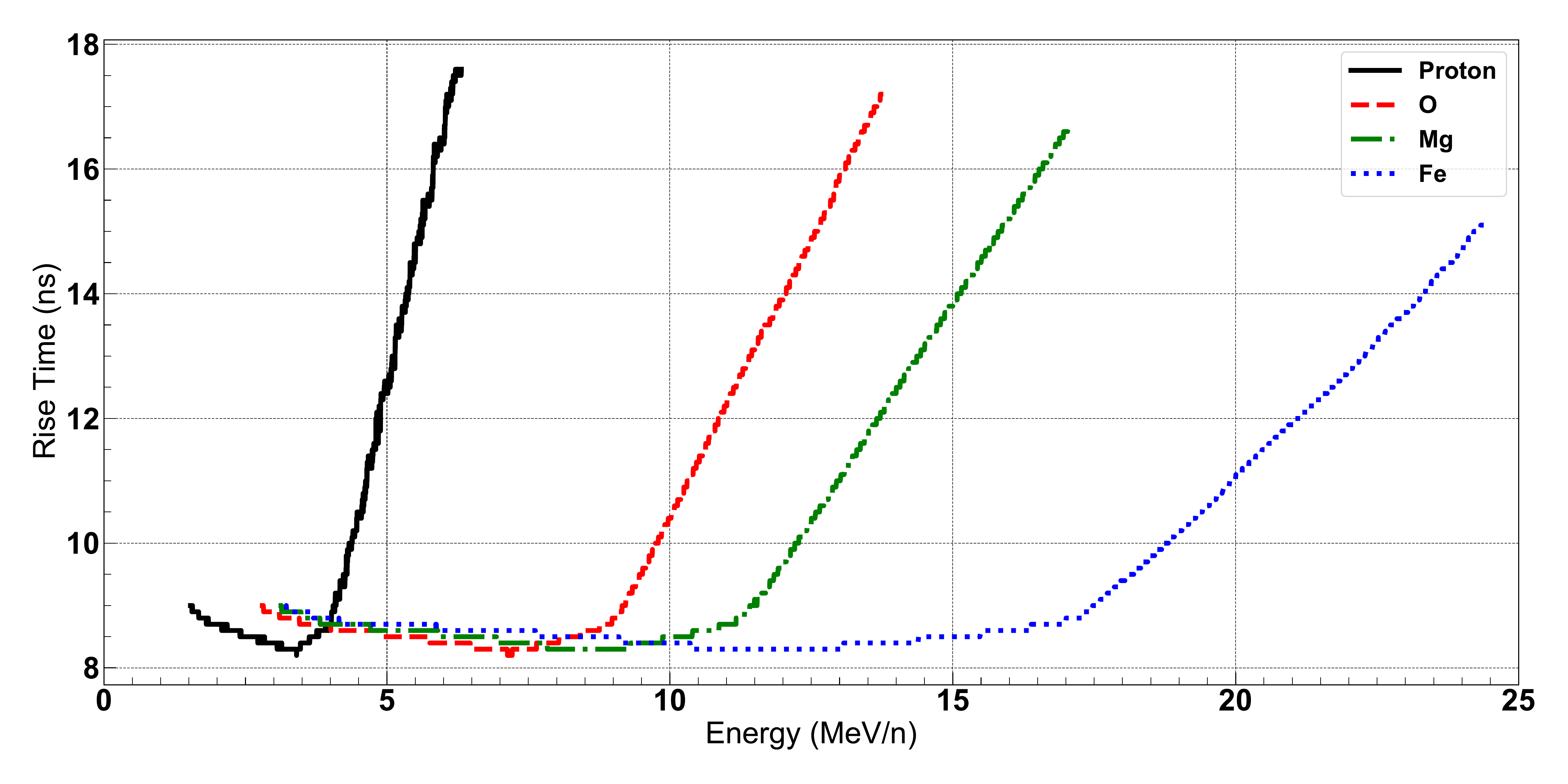}
    \caption{Rise time (ns) as a function of energy (MeV/n) in the first 300$\mu$m layer for stopping particles: proton (black full line), oxygen (red dashed line), magnesium (green dash-dotted line) and iron (blue dotted line) ions.}
    \label{fig:Rtime_s1}
\end{figure}

As can be seen in Fig. \ref{fig:Rtime_s1} each ion has a specific characteristic ``rise time vs energy dependence" which can be used to determine particle ID (Z and M) (see Section \ref{subsection:PSDAlgorithm}). However, there is a region (``V" shape area around 9 ns) where it is impossible to separate the curves for different ions with different energies as they overlap. This region corresponds to the case when a particle stops in the first part of the detector (1/3 of the total detector thickness for the default configuration used in the simulations). Since the hole (+) component of the current is slower than the electron component (-), the charge collection time and the rise time are only determined by the electron component of the signal which is constant for all ions. This sets the limits on the energy range for each ion that can be discriminated using the rise time and amplitude measurements (Section \ref{section:performance} provides examples for several ions). In this energy range, a rise time value is nearly indistinguishable for not only different particle types, but also for different energies within the same type.
One can also observe that the ``V" shape is broader for heavier ions due to their relatively shorter stopping range at the same energy (see Fig. \ref{fig:range}) which affects the energy range where these particles can be distinguished.

\subsubsection*{Pulse Amplitude}

\begin{figure}[!htp]
    \centering
    \includegraphics[width=0.85\textwidth]{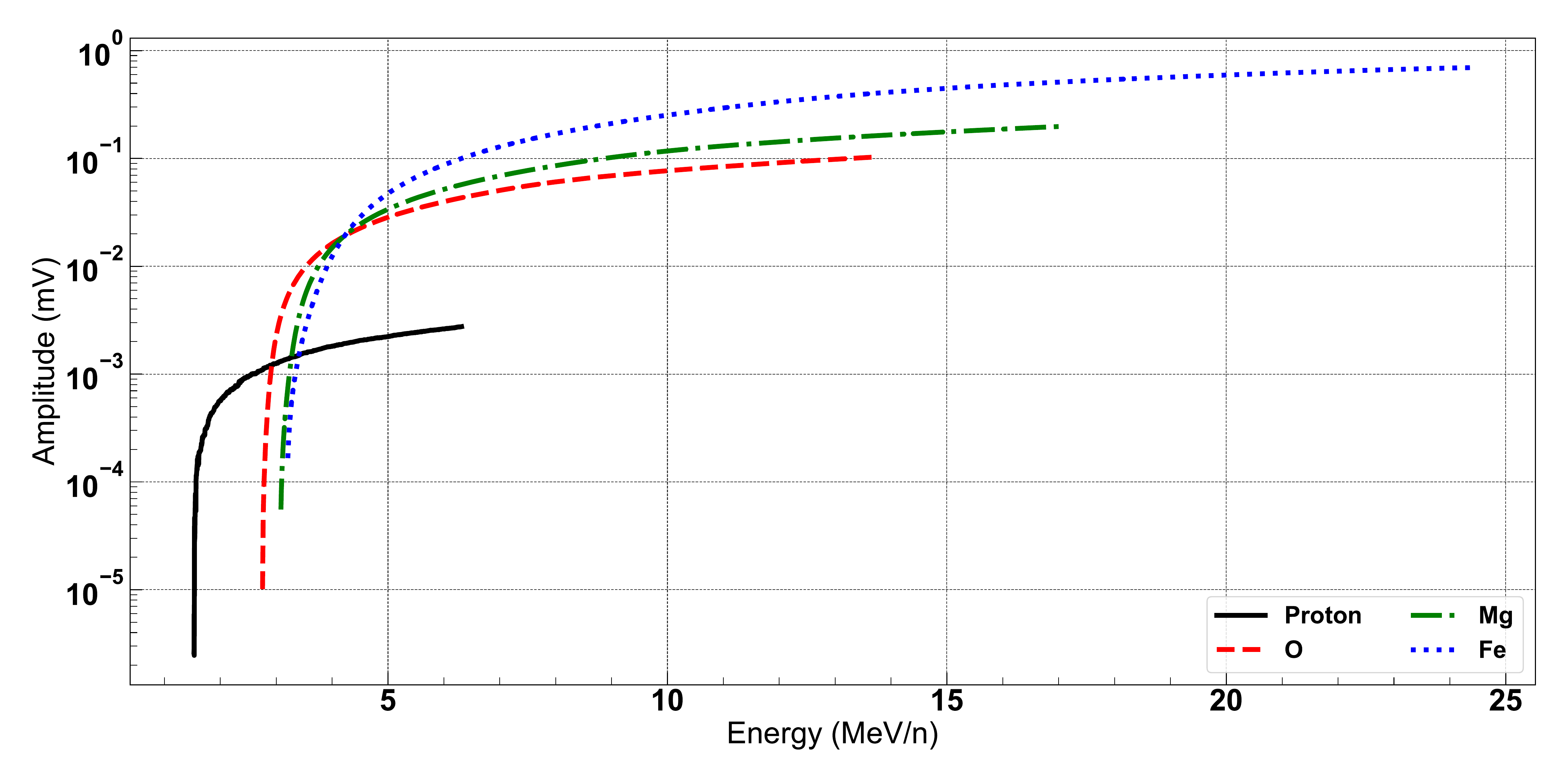}
    \caption{Maximum amplitude (mV) as a function of initial energy (MeV/n) for the particles stopping in the first layer. Proton (black full line), Oxygen (red dashed line), Magnesium (green dash-dotted line) and Fe (blue dotted line). }
    \label{fig:MaxA_s1}
\end{figure}

The amplitude (maximum voltage) of the pulse is proportional to the total charge collected, and thus to the total energy deposited in a detector layer. An example of this dependence is shown in Fig. \ref{fig:MaxA_s1}. The pulse amplitude over a range of energies differs for different ions (proton, oxygen, magnesium, and iron).

The possibility of accurate measurement of amplitude and its relevant dependence on the initial energy of the particle makes it the best key feature for energy estimation as long as the particle has already been identified.

\subsubsection*{Pulse Amplitude vs Rise Time}
As mentioned above, the two key characteristics in the stopping layer are the rise time and amplitude. Individually, these values can give partial information to characterize a particle, but when used together, they can identify both the particle type (Z and M) and its energy.

\par
\begin{figure}[!htp]
    \centering
    \includegraphics[width=1\textwidth]{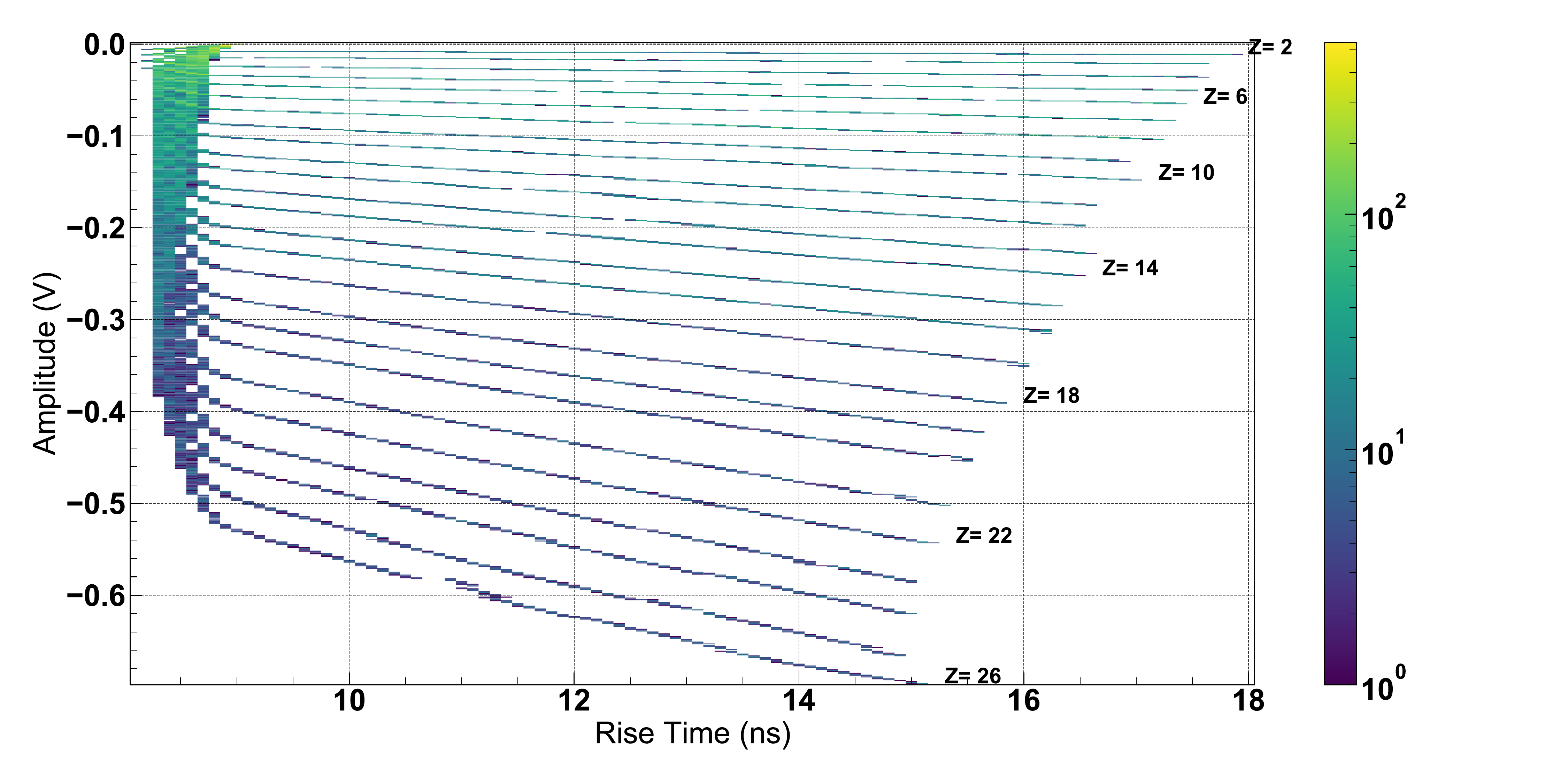}
    \caption{Maximum amplitude vs Rise time for He-Fe ions stopping in the detector (see Section \ref{subsection:range}). The color bar shows the number of events in each bin of this 2D histogram.}
    \label{fig:h_AvsT_all}
\end{figure}

\begin{figure}[!htp]
    \centering
    \includegraphics[width=1\textwidth]{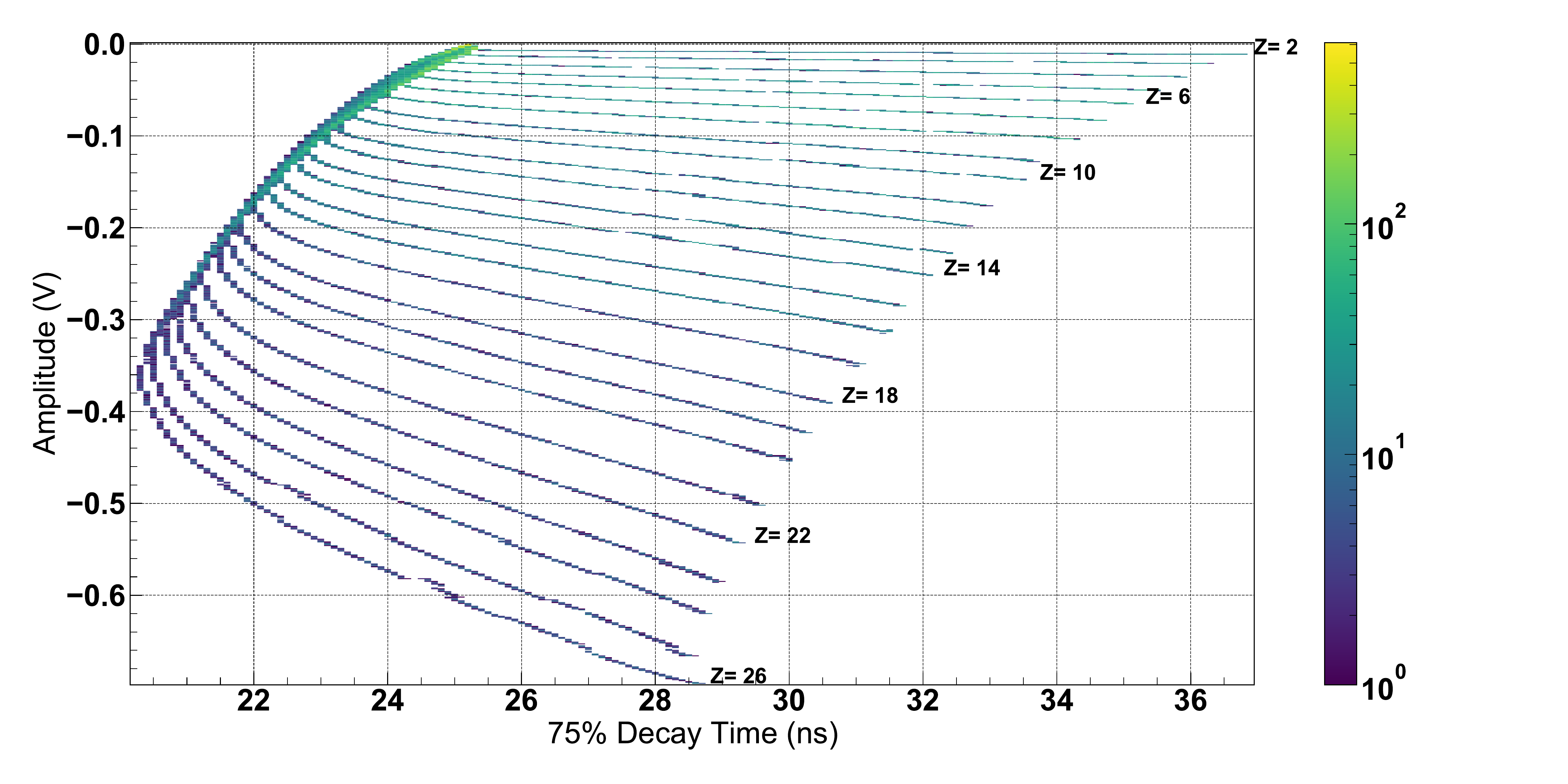}
    \caption{Maximum amplitude vs 75\% Decay time for He-Fe ions stopping in the detector. The color bar shows the number of events in each bin of this 2D histogram.}
    \label{fig:Avs75T_stopping}
\end{figure}

\begin{figure}[!htp]
    \centering
    \includegraphics[width=1\textwidth]{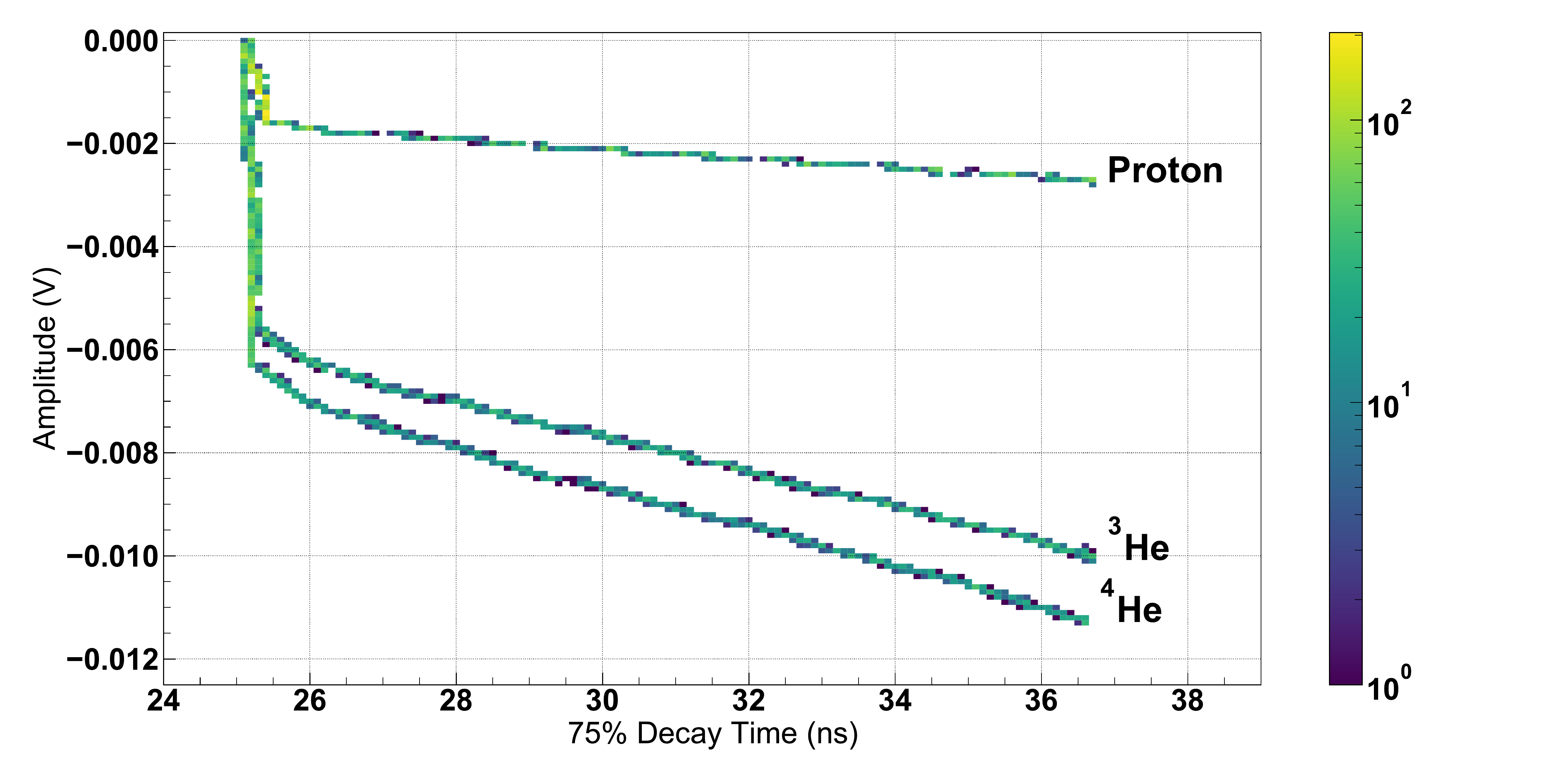}
    \caption{Maximum amplitude vs Rise time for Protons, \isotope{He}{3}, and \isotope{He}{4} ions stopping in the detector. The color bar shows the number of events in each bin of this 2D histogram.}
    \label{fig:Avs75T_P}
\end{figure}

 Due to the specifics of the read-out electronics (Section \ref{subsection:amps}), the time when the amplitude of the pulse decreases to 75\% maximum value ("75\% decay time", see Fig. \ref{fig:ex_lowgain}) is more effective than the rise time itself. In both cases (see Fig. \ref{fig:h_AvsT_all} and Fig. \ref{fig:Avs75T_stopping}), the curves are unique for each ion and thus can be used for particle identification; particle energy is then determined using only the maximum amplitude. The details regarding the performance of the method are presented in Section \ref{section:performance}.
It should be mentioned that in addition to charge and mass determination, different isotopes can also be discriminated using the PSD technique. As an example we show He isotope separation achieved by the PSD technique in Fig. \ref{fig:Avs75T_P}. As can be seen in the Figure, the two main isotopes of helium (\isotope{He}{3} and \isotope{He}{4}) are well distinguished from each other as well as from the protons. The same method can be applied to distinguish isotopes with Z up to 26. 

\section{The Detection System}
\label{S:instrument}
\subsection {Silicon Detectors}
\label{subsection:silicon_detectors}
The Si detectors used in the simulations (MSD040 \cite{MSD040}) are  manufactured by Micron Semiconductor Ltd (United Kingdom) with an active area diameter of 40 mm and a thickness of 300 $\mu$m. Some key characteristics of MSD040 detectors used in these studies are given in Table \ref{table:MSD040}.
\begin{table}[!htp]

    \centering
    \begin{tabular}{c|c}
    \hline
       Active area diameter & $40 mm$ \\
         \hline
        Thickness  &   $300 {\mu}m$ \\
         \hline
         Typical Full Depletion (FD)  & $<60V$  \\
         \hline
         Total Leakage Current (at FD +10V) &  $<10nA$ \\
         \hline
        Capacitance (FD) & 40 $pF/cm^2$ \\
         \hline
        Resistivity & (3 – 10) $k{\Omega}{\times}cm$ \\
      
    \end{tabular}
    \caption{Main operational characteristics of the MSD040 detectors}
    \label{table:MSD040}
\end{table}

Operational characteristics such as detector thickness and bias voltage (110V) were chosen in order to get a large effective area (geometry factor) and to cover a wide energy range while providing stable and efficient performance of the read-out electronics (both analog and digital) in regard to the detector capacitance, the signal amplitude, and length.

\subsection {Read-Out Electronics (Amplifiers)} 
\label{subsection:amps}
The key principles and components for the amplifiers used in the current studies were adapted from the multipurpose read-out board designed at the University of Kansas \cite{minafra2017}. The duration of the useful output signal in the chosen configuration is less than 50 ns (Fig. \ref{fig:ex_lowgain}).
While the signal shape is completely simulated, to simplify the analysis the noise is assumed as $\sim 1mV$. In fact (from prior experience), the noise on the real front-end electronics is expected to be lower. 

\subsection {The PSEC4 Sampler} 
\label{subsection:psec}
To get the best results from the digital PSD method, the signals from the detector should be digitized with high time and amplitude resolutions. For this purpose, the PSEC4 chip \cite{oberla201415} \footnote{Another possible candidate for a sampler is the SAMPIC (SAMpler for PICosecond time pick-off) chip \cite{delagnes2014sampic} which has comparable characteristics as PSEC4.} is used currently in the AGILE project and its main characteristics are included in the performance simulations (Table \ref{table:PSEC4}).

\begin{table}[!htp]

    \centering
    \begin{tabular}{c|c}
    \hline
       Switched Capacitors Array (SCA) Depth & 256 samples \\
         \hline
        Sampling Rate  &   (4-15) GSa/s \\
         \hline
         ADC DC Dynamic range  & 10.5 bits  \\
         \hline
         Bandwidth &  1.5 GHz \\
         \hline

    \end{tabular}
    \caption{PSEC4 Sampler: main characteristics}
    \label{table:PSEC4}
\end{table}

\section{Simulated Performance}
\label{section:performance}
As discussed in Section \ref{subsection:simShort}, detailed simulations were used to obtain the relevant characteristics presented in Section \ref{ss:signalCharacteristics} and used in the PSD method. The simulation also provides systematic uncertainties on these values, arising for example, from noise in the electronic chain. This section first describes the main restriction on particle identification that limits the acceptance range and the performance of energy estimation of our method.

\subsection{Energy Acceptance for PSD method}
\label{energy_accepatnce}
\begin{figure}[!htp]
    \centering
    \includegraphics[width=1\textwidth]{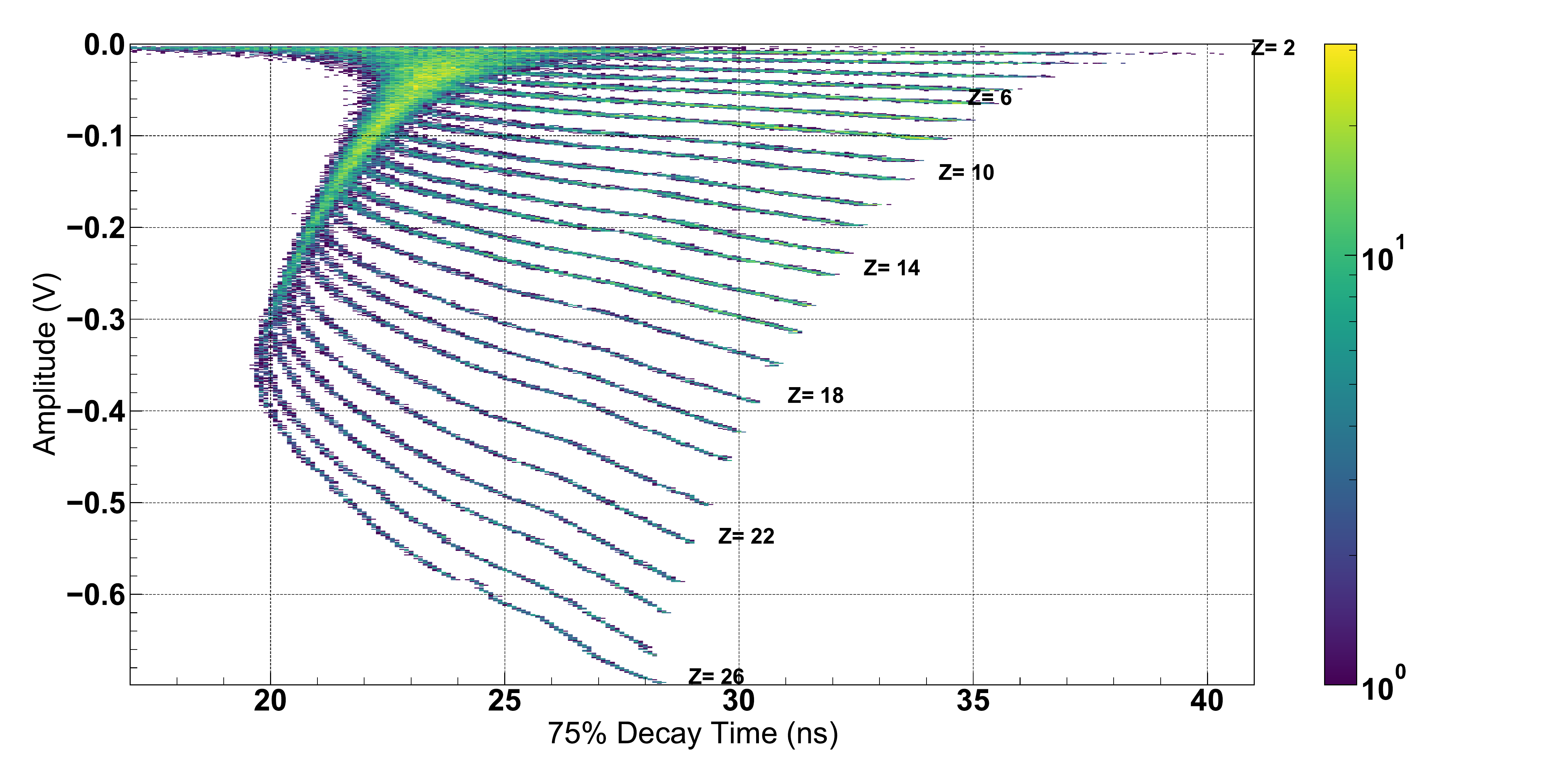}
    \caption{Maximum amplitude vs 75\% decay time for stopping particles when noise and jitter are simulated. The color bar shows the number of events in each bin of this 2D histogram.} 
    \label{fig:Avs75T_stopping_jitter}
\end{figure}

While Fig. \ref{fig:Avs75T_stopping} shows the theoretical (only statistical fluctuations are included, see Section \ref{subsection:uncertainties}) performance of the ion discrimination method, it does not take into account the electronic noise affecting amplitude measurements and the expected jitter for the timing measurements. Fig. \ref{fig:Avs75T_stopping_jitter} shows how the maximum amplitude and 75\% decay time are affected when when those perturbations are taken into account. As expected low amplitude signals are drastically impacted by jitter which prohibits discrimination of most light ions (Z$<$4) in the regions where the curves overlap. In order to address this the amplifier of the instrument needs to be adapted.

Fig. \ref{fig:Avs75T_stopping_jitter} reveals a region where curves for different ions at low energies overlap. The overlap is mainly due to the fact that all charge carriers are created in the first 1/3 of the detector (a ``V shape" region in the timing measurements). Trying to determine a particle ID in this region of overlap leads to a high level of misidentification. \par
\begin{figure}[!htp]
    \centering
    \includegraphics[width=0.9\textwidth]{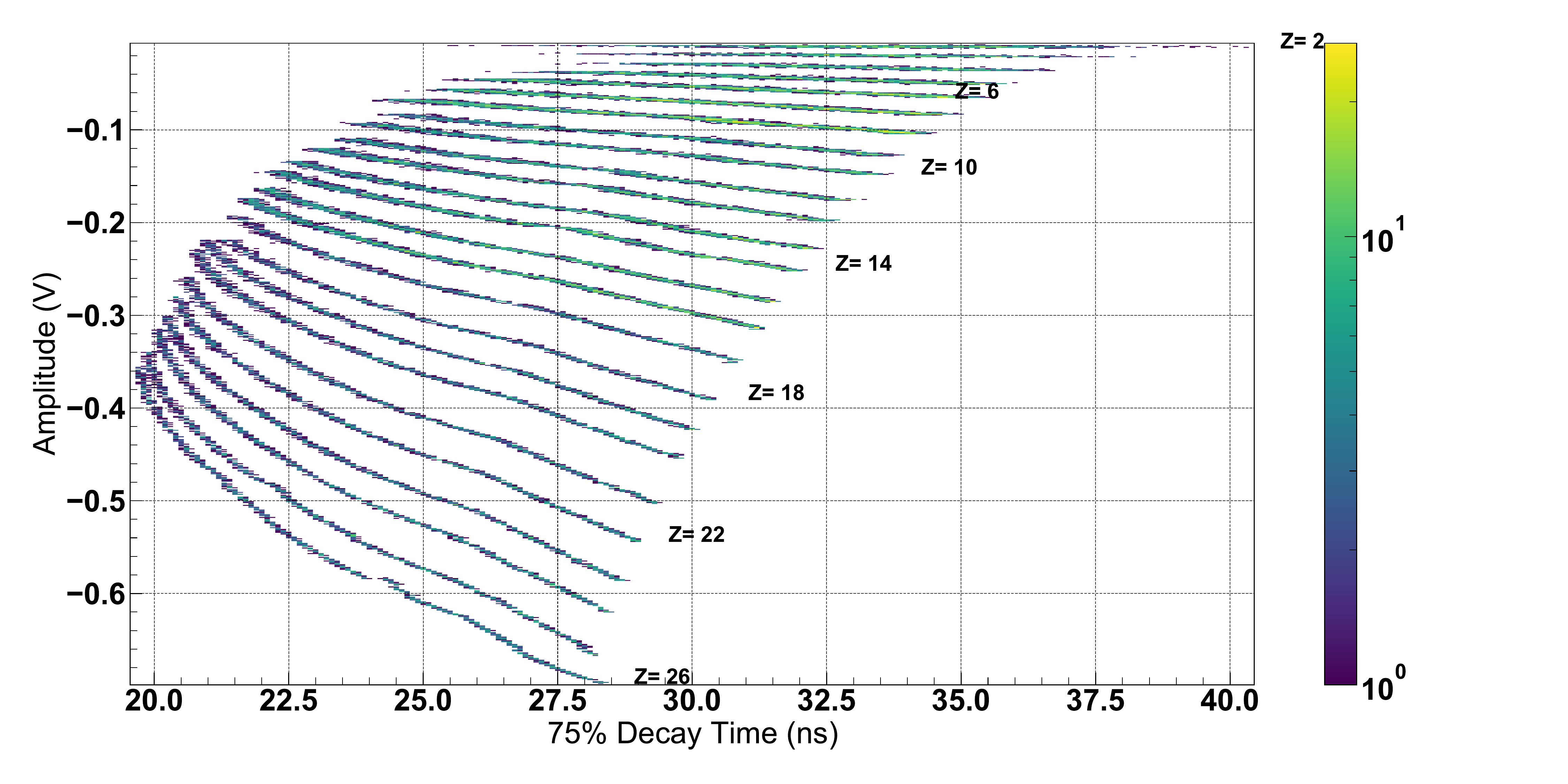}
    \caption{Maximum amplitude vs 75\% decay time for stopping particles after energy restriction in discrimination acceptance. This plot is used to discriminate the particle ID of the measured event.}
    \label{fig:IDdiscrimination}
\end{figure}

To prevent this misidentification, a special selection was made on a domain of energy range different for each ion. The events from the overlapping region are considered to be ``bad" pulses and are discarded. Fig. \ref{fig:IDdiscrimination} shows how this selection affects the ID discrimination. For example, signals from Oxygen were selected above 59 mV (absolute value), discarding all Oxygen events with a maximum amplitude lower than this value, which corresponds to the energy of 16 MeV/n (e.g. for Mg these values are 109 mV and 20 MeV/n respectively). Thus each layer has a restricted domain of energy for each ion. \par
However, in order to determine the spread of the overlapping region, simulated data for all possible ions are considered as the signals from heavier ions affect the energy selection for the lighter ones.

\begin{figure}[!htp]
    \centering
    \includegraphics[width=1\textwidth]{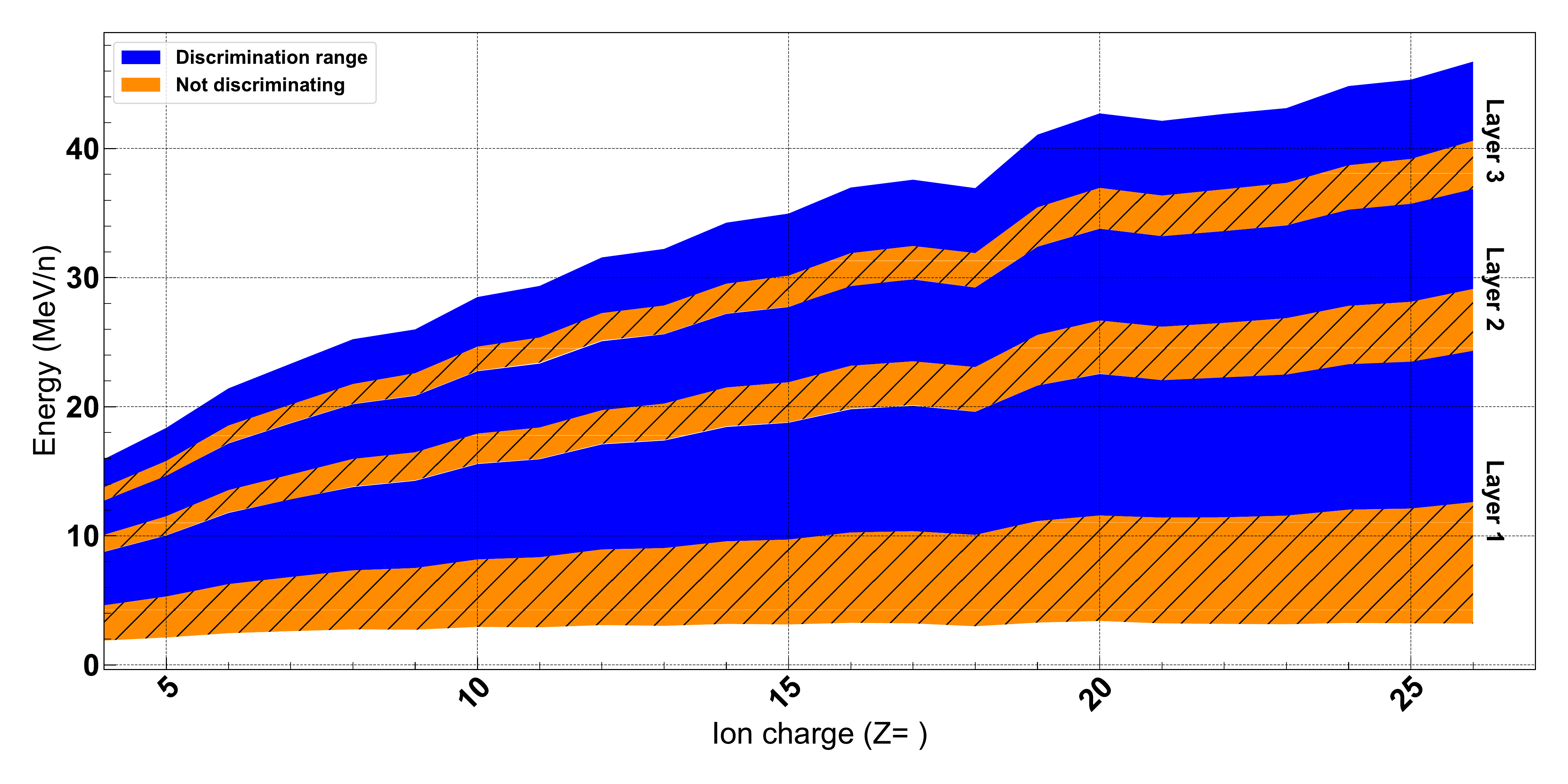}
    \caption{Energy vs Charge (Z). Plain blue regions are the domain where the method can identify the particles. The hatched orange regions are the domain where the discrimination is not possible.}
    \label{fig:Eacceptance}
\end{figure}

This selection modifies the acceptance of the method for each specific ion since the energy range where discrimination is possible depends on the type of particle. Fig. \ref{fig:Eacceptance} shows these ranges of good acceptance for each ion. Orange areas represent particles stopping in a layer with an energy outside of the range of acceptance, the type of those particles cannot be determined.

\begin{figure}[!htp]
    \centering
    \includegraphics[width=1\textwidth]{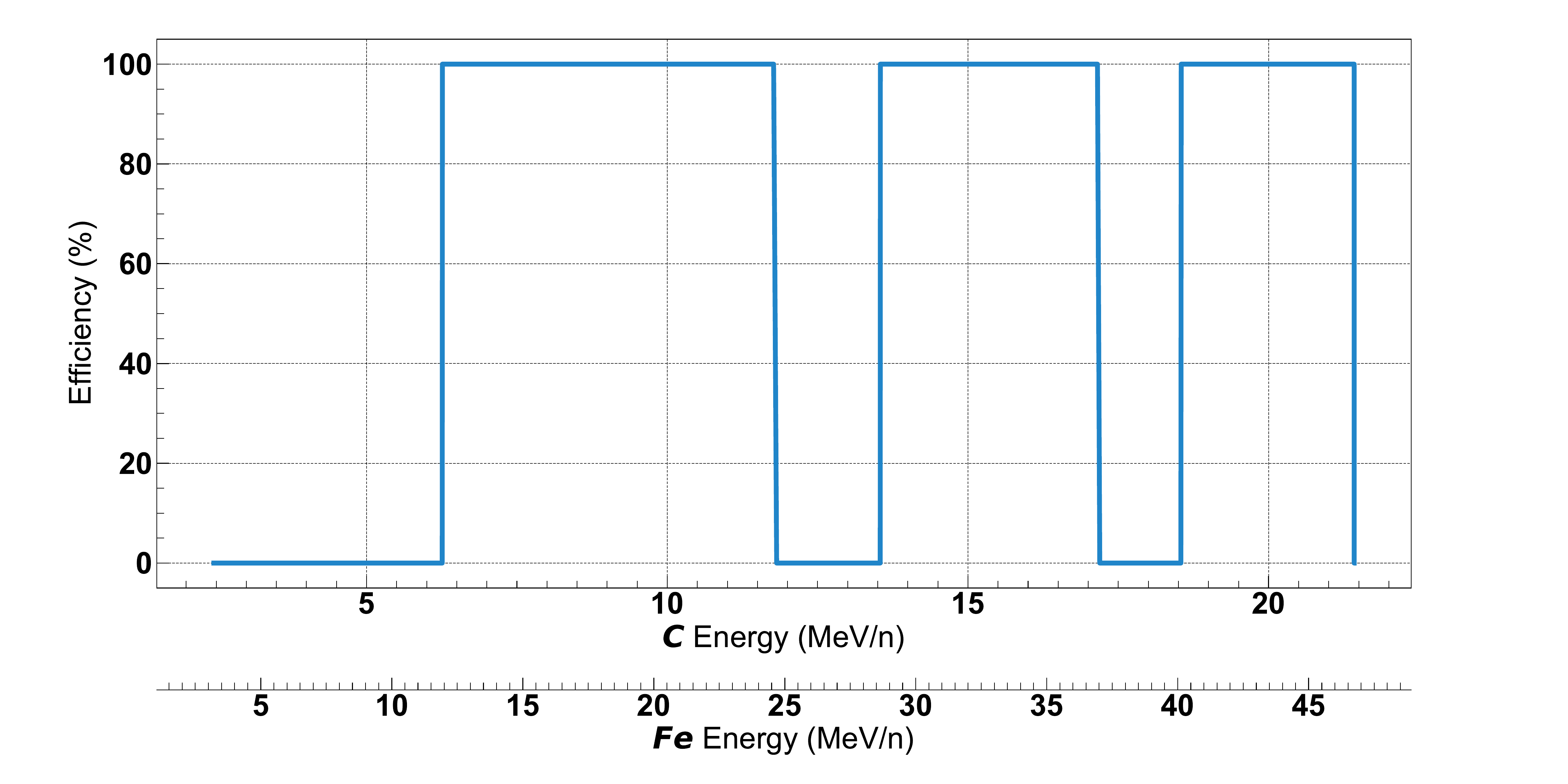}
    \caption{Efficiency of the particle ID determination for two ions (C and Fe).}
    \label{fig:efficiency}
\end{figure}

In the energy ranges where identification is possible, its efficiency is close to 100\%. An example is shown in Fig. \ref{fig:efficiency} for C and Fe ions. The abrupt change of efficiency between 100\% and 0\% is, in reality, smoother.

This acceptance was determined from the simulation of the hardware and will be calibrated and tested. The fact that we perform measurements only on discrete energy domains which are particle dependent is not a drawback for the measurements AGILE plans to perform since particle spectra can be obtained with non-overlapping differential energy bins.

\subsection{Energy Measurement}
\begin{figure}[!htp]
    \centering
    \includegraphics[width=1\textwidth]{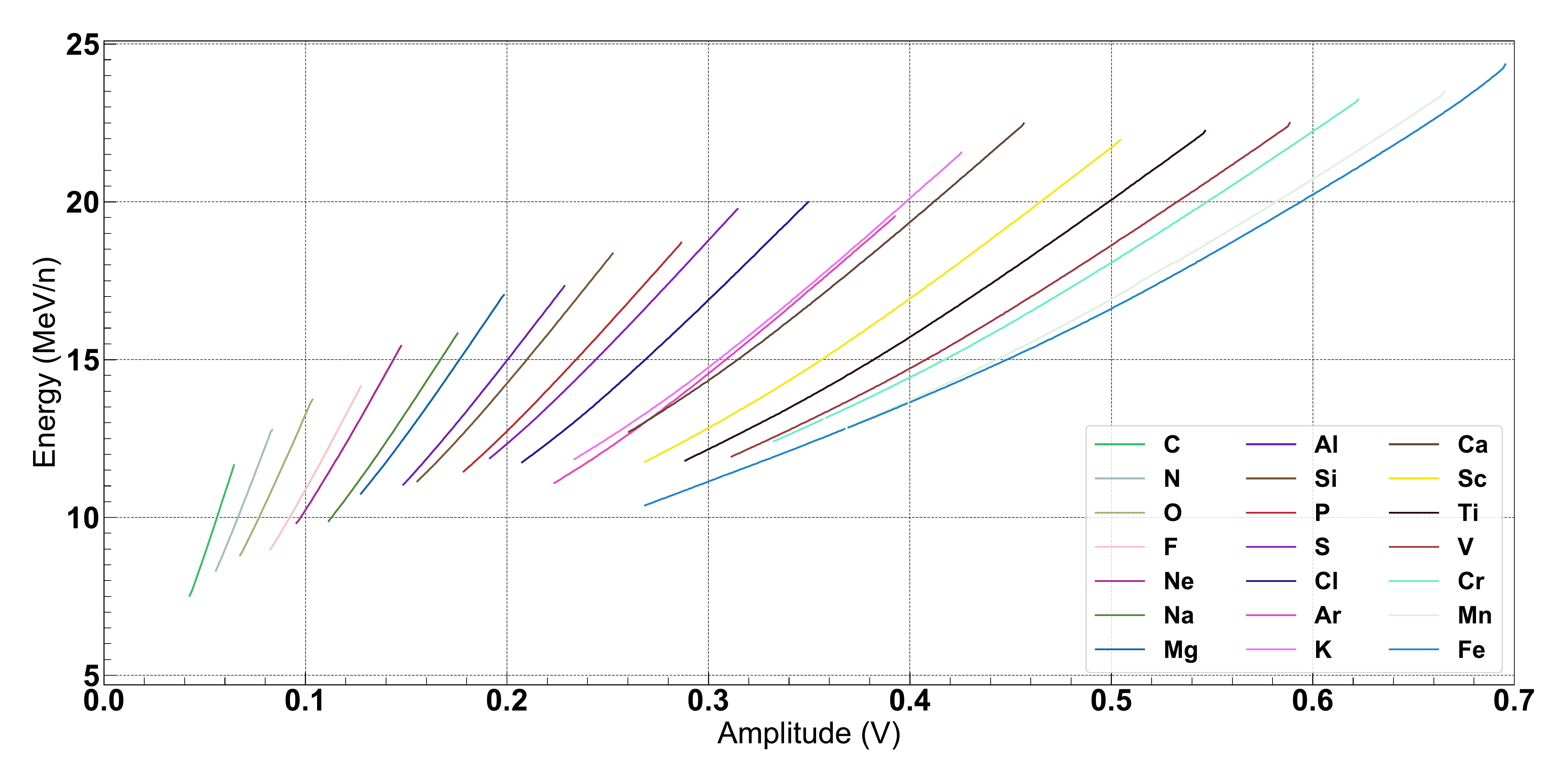}
    \caption{Estimated energy as a function of the measured maximum amplitude in layer 1 when the determination of particle ID is possible. Line widths indicate errors on energy. Lines from left to right are for C to Fe.}
    \label{fig:Ediscrimination}
\end{figure}

 To measure particle energies, the PSD method relies on using the amplitude of the signal. Fig. \ref{fig:Ediscrimination} shows the estimated incident energy of the particle as a function of the simulated maximum amplitude for each ion.

%
%
In the following, the systematics that affect the performance of the method are discussed.
\subsubsection{Major Sources of Uncertainties }
\label{subsection:uncertainties}
One of the main sources of uncertainties in the pulse shape, and therefore the values of its key characteristics (time and amplitude), originates from the statistical nature of charge formation inside the detector volume. There are two factors contributing to this uncertainty:
\begin{itemize}
    \item The statistical behaviour of the energy deposition along the track of the incoming particle, so-called ``Energy Straggling" \cite{landau1944physics, vavilov1957ionization};
    \item The statistical fluctuation of the number of charge carriers (electron-hole pairs) produced in the detector medium \cite{fano1947ionization}.
\end{itemize}

This uncertainty cannot be corrected since it is intrinsic to the physical processes of ions interacting with the detector medium (Si). \par
 Electronics noise depends on the hardware and on its specific design. Temperature effects on both the signal formation and the electronics performance can be taken into account and corrected. Last but not least, the variation of the particle angle of incidence needs to be taken into account, since AGILE will be sensitive to particles originating from different directions. \par

Based on these factors, the uncertainty on the measured particle energy can be determined and therefore the energy resolution of the proposed detection system can be estimated (as 1$\sigma$ of the energy distribution), as described in the following subsections.

 {\bf{Statistical fluctuations}} \par
\begin{figure}[H]
    \begin{subfigure}{1.0\textwidth}
        \centering
        \includegraphics[width=1\textwidth]{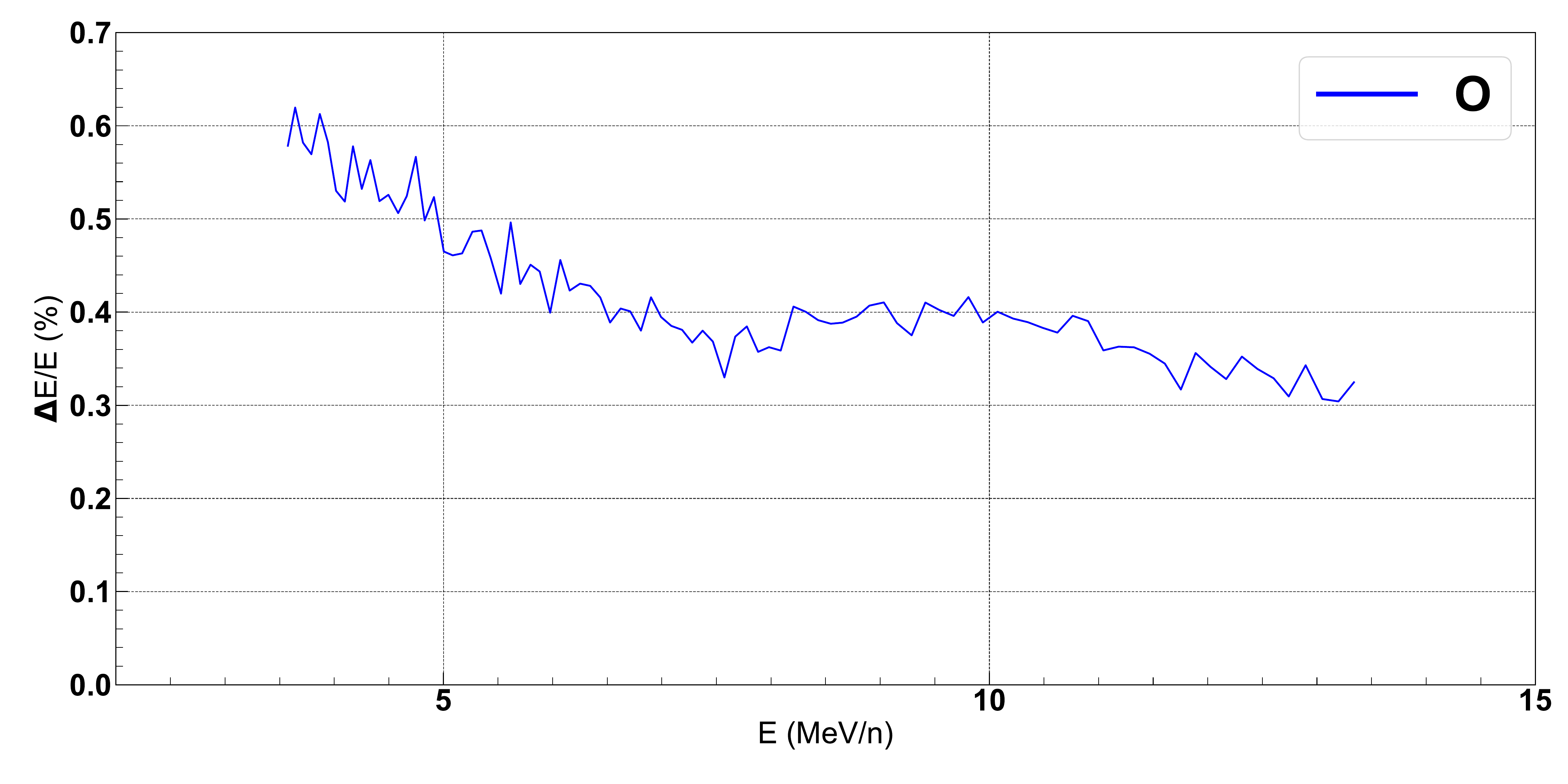}
    \end{subfigure}
    \begin{subfigure}{1.0\textwidth}
        \centering
        \includegraphics[width=1\textwidth]{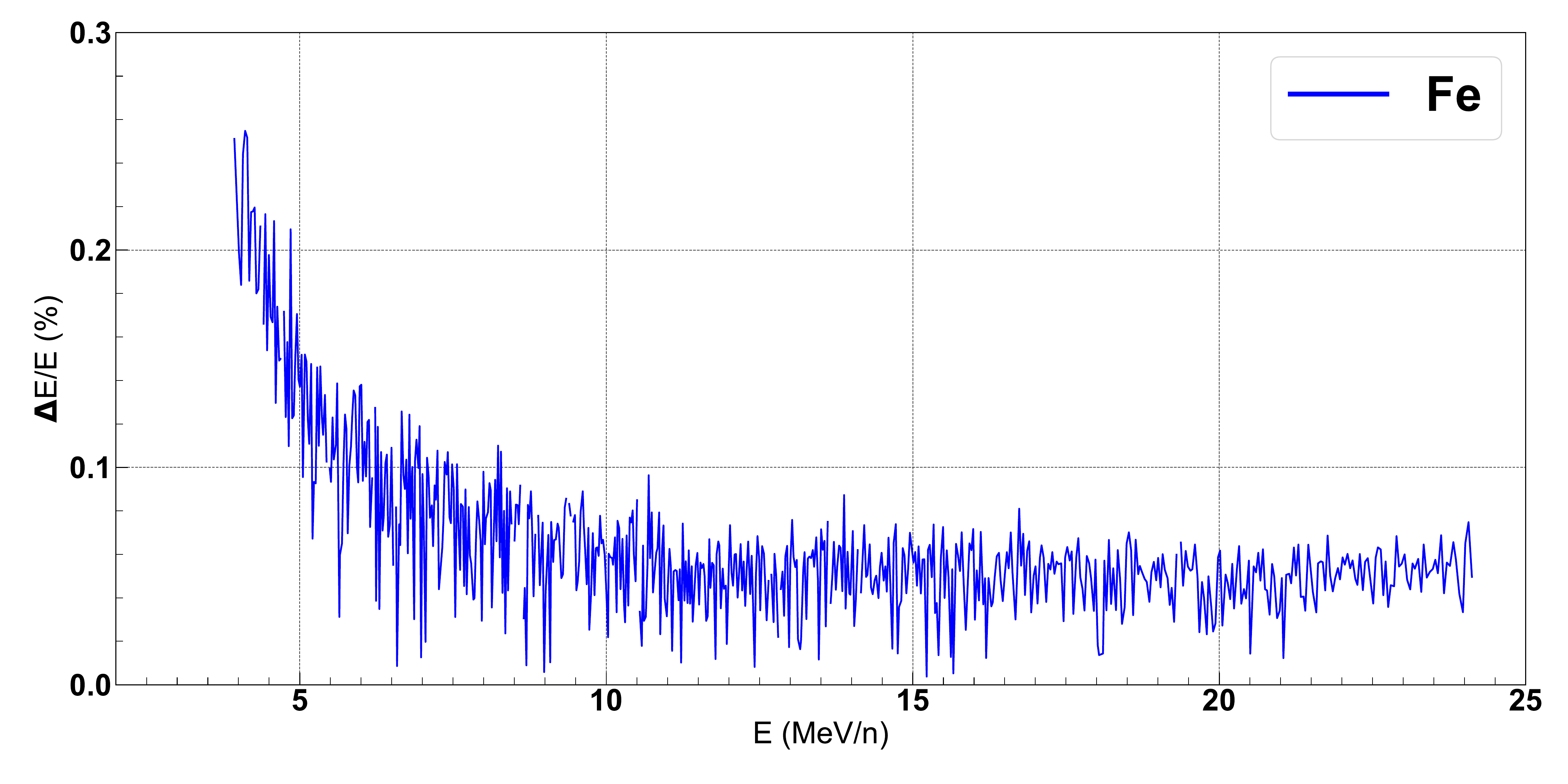}
        
    \end{subfigure}
    \caption{Relative energy resolution due to statistical fluctuations of the energy deposition and the number of charge carriers. Top: oxygen, bottom: iron}
    \label{fig:resolution}
\end{figure}

The intrinsic fluctuation of energy deposition and carrier numbers leads to a non-negligible uncertainty that affects the maximum amplitude measured and therefore the actual energy resolution for each ion (Fig. \ref{fig:resolution}).

The energy deposition grows with the particle mass leading to a higher number of charge carriers in the silicon layer. This translates into a few orders of magnitude difference in the amplitude between light and heavy ions (Fig. \ref{fig:MaxA_s1}) and leads to a better relative energy resolution ($\Delta$E/E) for heavier ions. Lighter ions, such as oxygen, show an energy resolution of less than a percent while for heavier ions, such as iron, we get about $0.1\%$ which is very good compared to a required value of the order of 10\% for AGILE project. \par

{\bf{Electronics noise}} \par
\begin{figure}[H]
    \begin{subfigure}{1\textwidth}
        \centering
        \includegraphics[width=1\textwidth]{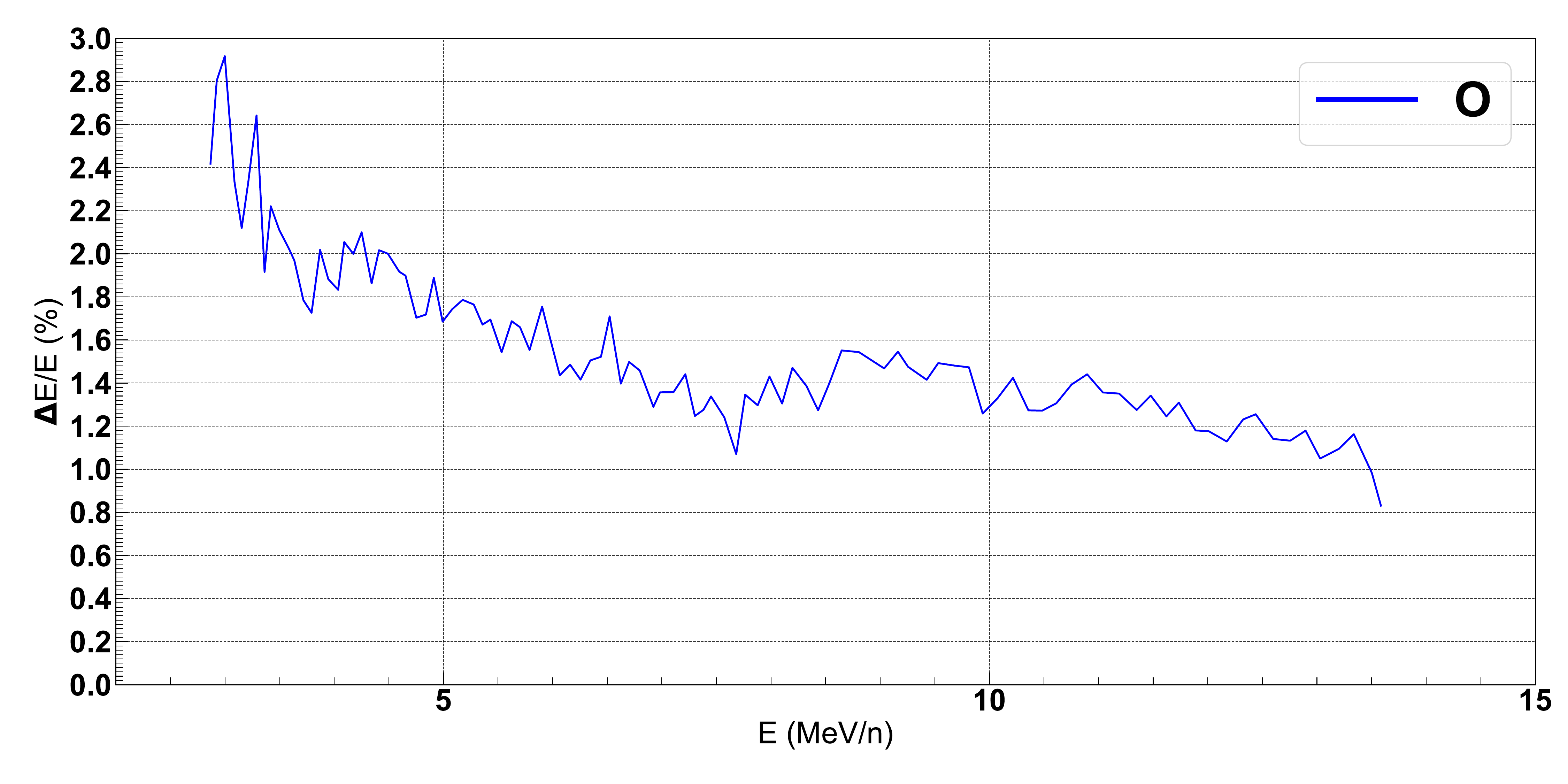}
        
    \end{subfigure}
    \begin{subfigure}{1\textwidth}
        \centering
        \includegraphics[width=1\textwidth]{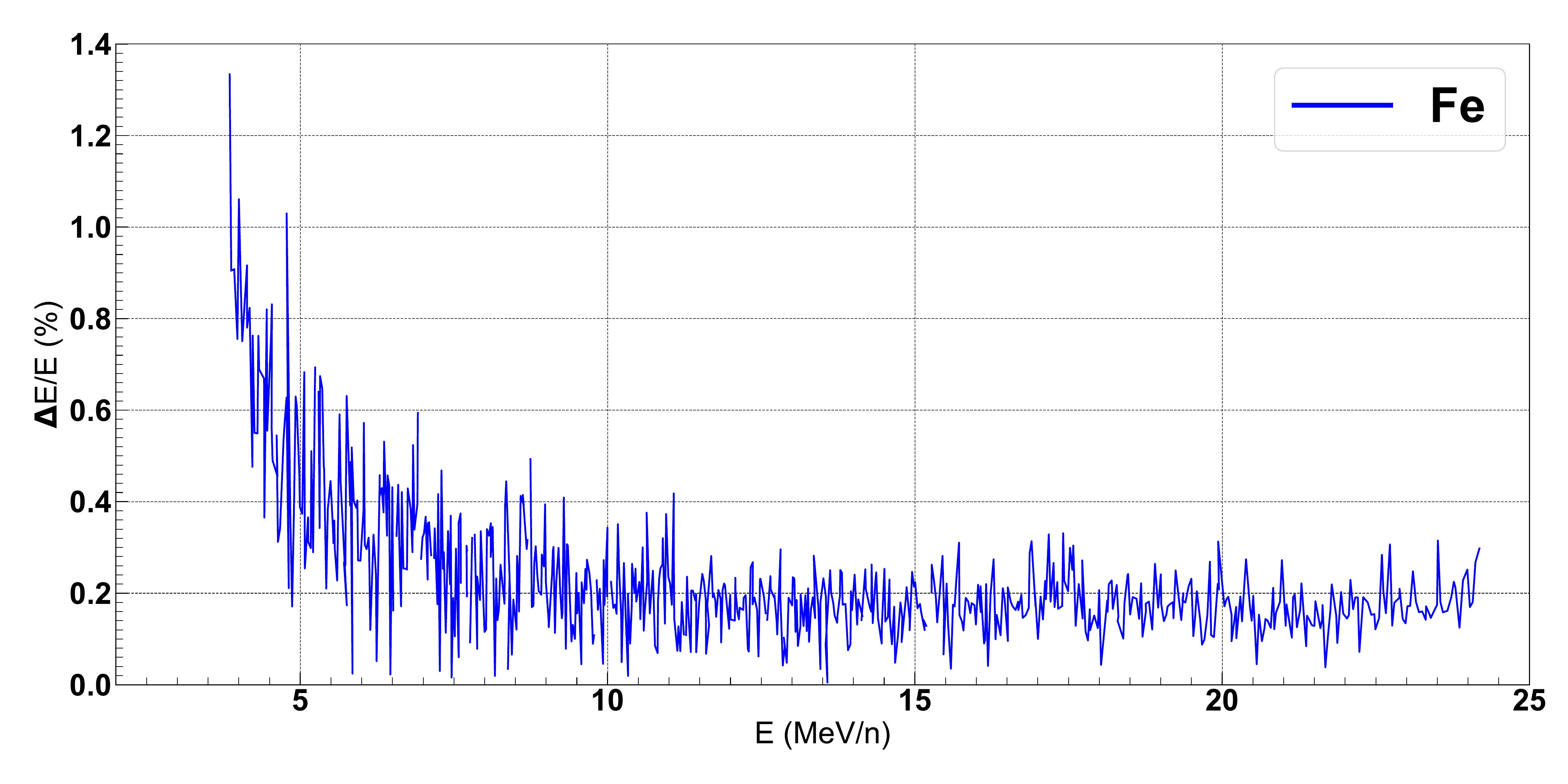}

    \end{subfigure}
    \caption{Energy resolution due to statistical fluctuations of the energy deposition and the number of charge carriers and electronics noise. Top: oxygen, bottom: iron.}
    \label{fig:resolution_noise}
\end{figure}

A preliminary estimation of the expected electronics noise leads to an approximate value of 1 mV. The impact of this noise is more significant for lighter ions due to their lower maximum amplitude. This electronics noise is added to the statistical fluctuations discussed above and the total energy resolution is presented in Fig. \ref{fig:resolution_noise} for O and Fe, as an example.

The impact of electronic noise is quite important for light ions, bringing the energy resolution for oxygen to $\sim 2\%$ while the resolution for iron remains below $1\%$, far better than the requirement of AGILE. \par

  {\bf {Incident angle variation}}\par
\begin{figure}[H]
    \centering
    \includegraphics[width=0.85\textwidth]{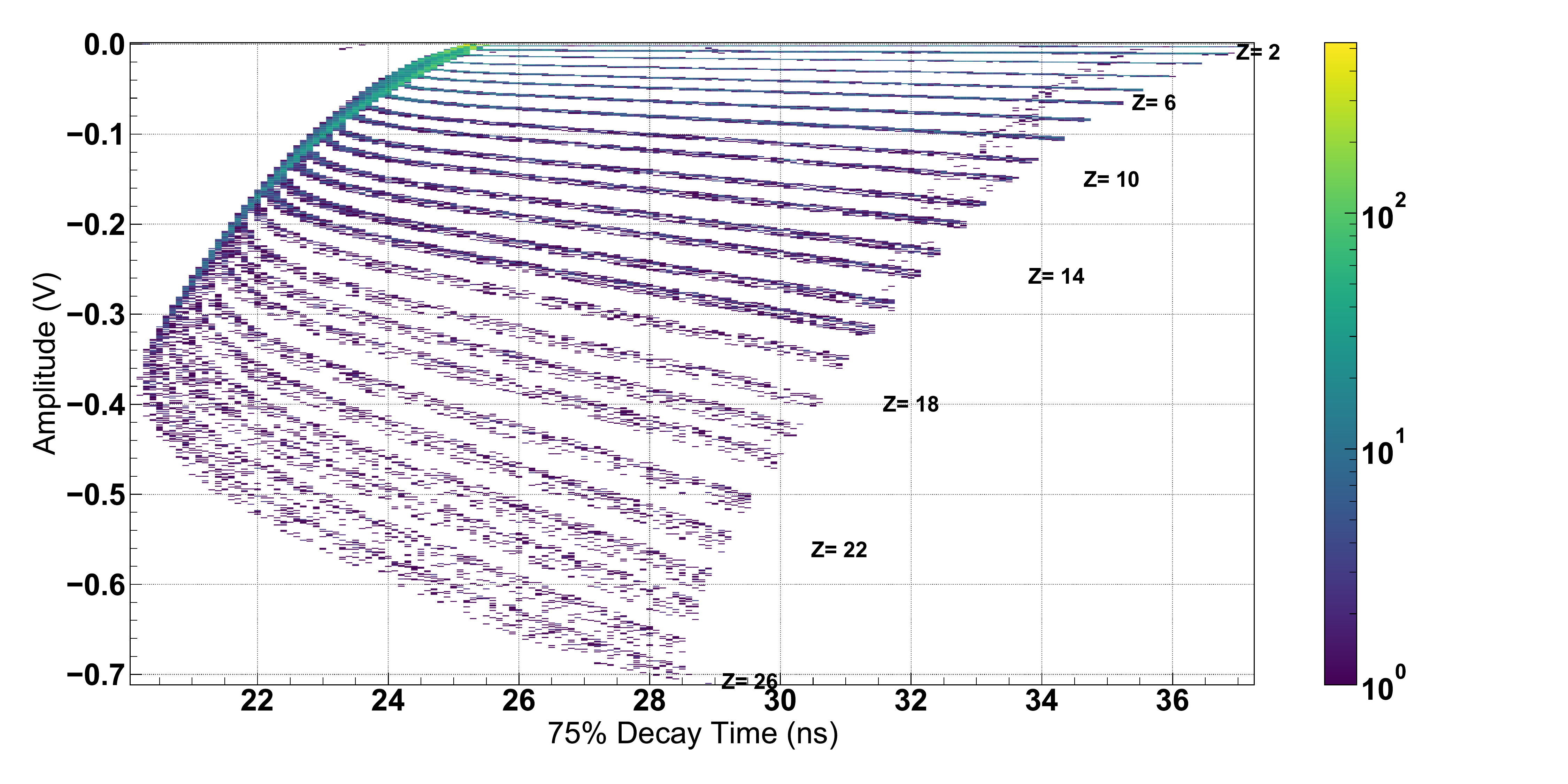}
    \caption{Maximum amplitude vs 75\% decay time for stopping particles for a field of view of 40 degrees in layer 1.}
    \label{fig:20deg}
\end{figure}
\begin{figure}[H]
    \centering
    \includegraphics[width=0.85\textwidth]{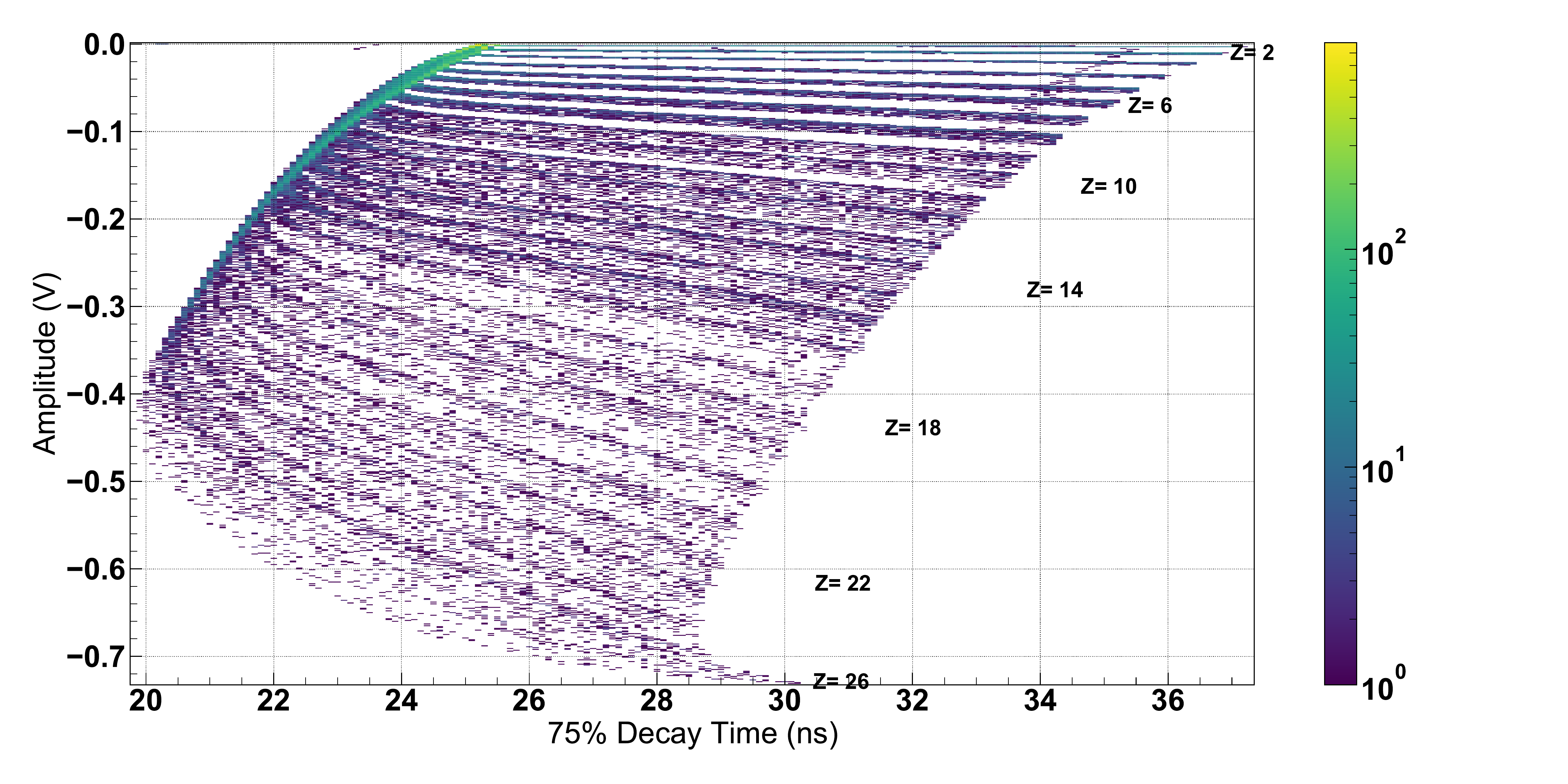}
    \caption{Maximum amplitude vs 75\% decay time for stopping particles for a field of view of 70 degrees in layer 1.}
    \label{fig:35deg}
\end{figure}

The spread in direction of the incoming particles was considered in the simulation, since the uncertainty caused by various angles of incidence is important.
Fig. \ref{fig:20deg} and Fig. \ref{fig:35deg} show the effect of incident angle variation on the performance of the discrimination method for fields of view of $40^{\circ}$ ($20^{\circ}$ of half-angle) and $70^{\circ}$ ($35^{\circ}$ of half-angle)\footnote{The cut on energy (Section \ref{energy_accepatnce}) was not applied in these figures.}.


For AGILE a field of $40^{\circ}$ is acceptable for all ions as no overlaps are observed. A wider field of view will endanger the capability of discrimination of heavy ions. The spread on the amplitude and timing values for lighter ions is less impacted than for heavier ones. 

{\bf {Temperature variations}} \par
\label{subsection:temperature} \par
In order to apply the method of particle ID and energy measurements on board a spacecraft, the effects of temperature on its performance need to be studied. In particular, the dependence of the following aspects on temperature changes the shape of the output signal and thus, the values of the key characteristics:
\begin{itemize}
    \item  mobility of charge carriers (electrons and holes) in the detector medium (Si);
    \item energy per electron-hole pair in silicon; 
    \item read-out electronics (amplifier) performance.
\end{itemize}

\begin{figure}[H]
    
        \centering
        \includegraphics[width=0.65\textwidth]{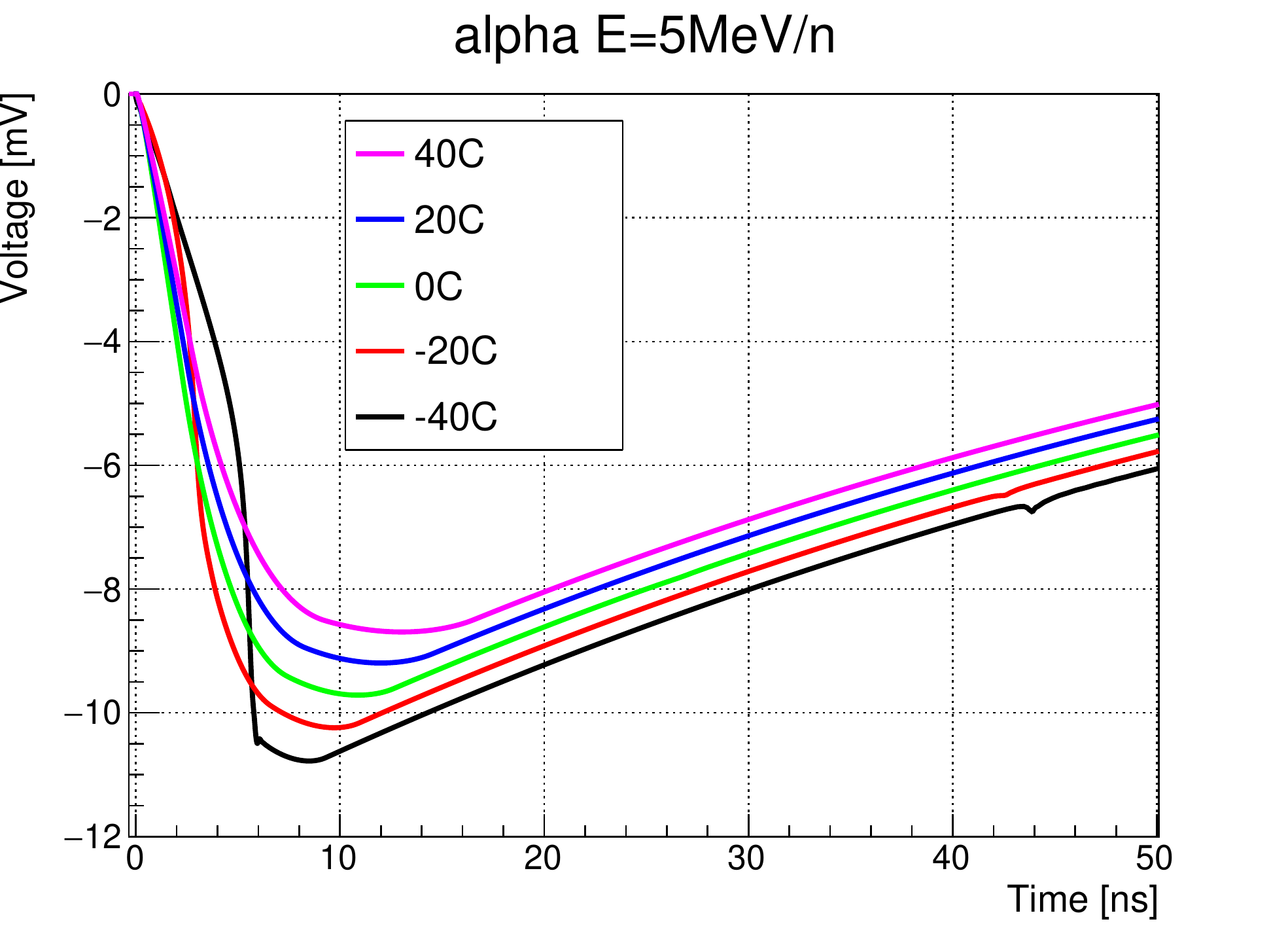} 
        \caption{ Dependence of the detector output signal on temperature. The long exponential decay part of the signal determined by RC time constant of the amplifier is not shown and only the ``useful" part of the signal $\sim$ 50 ns is presented. Because of the specifics of the signal formation and the read-out electronics performance the pulse shape at $-40^{\circ}$C differs from the others, however as shown below its main characteristics can be "corrected" based on the temperature.}
        \label{fig:temperature_voltage}

\end{figure}

Fig. \ref{fig:temperature_voltage} shows the results of the detector simulation for $\alpha$-particles with an energy of 5 MeV/n that stop in the first detector layer at various temperatures in the range of $-40^{\circ}$C to $+40^{\circ}$C. The simulations take into account the temperature effects on both stages, at the level of detector response (i(t)) as well as at the read-out electronics.

\begin{figure}[H]
    \begin{subfigure}{.5\textwidth}
        \centering
        \includegraphics[width=1\textwidth]{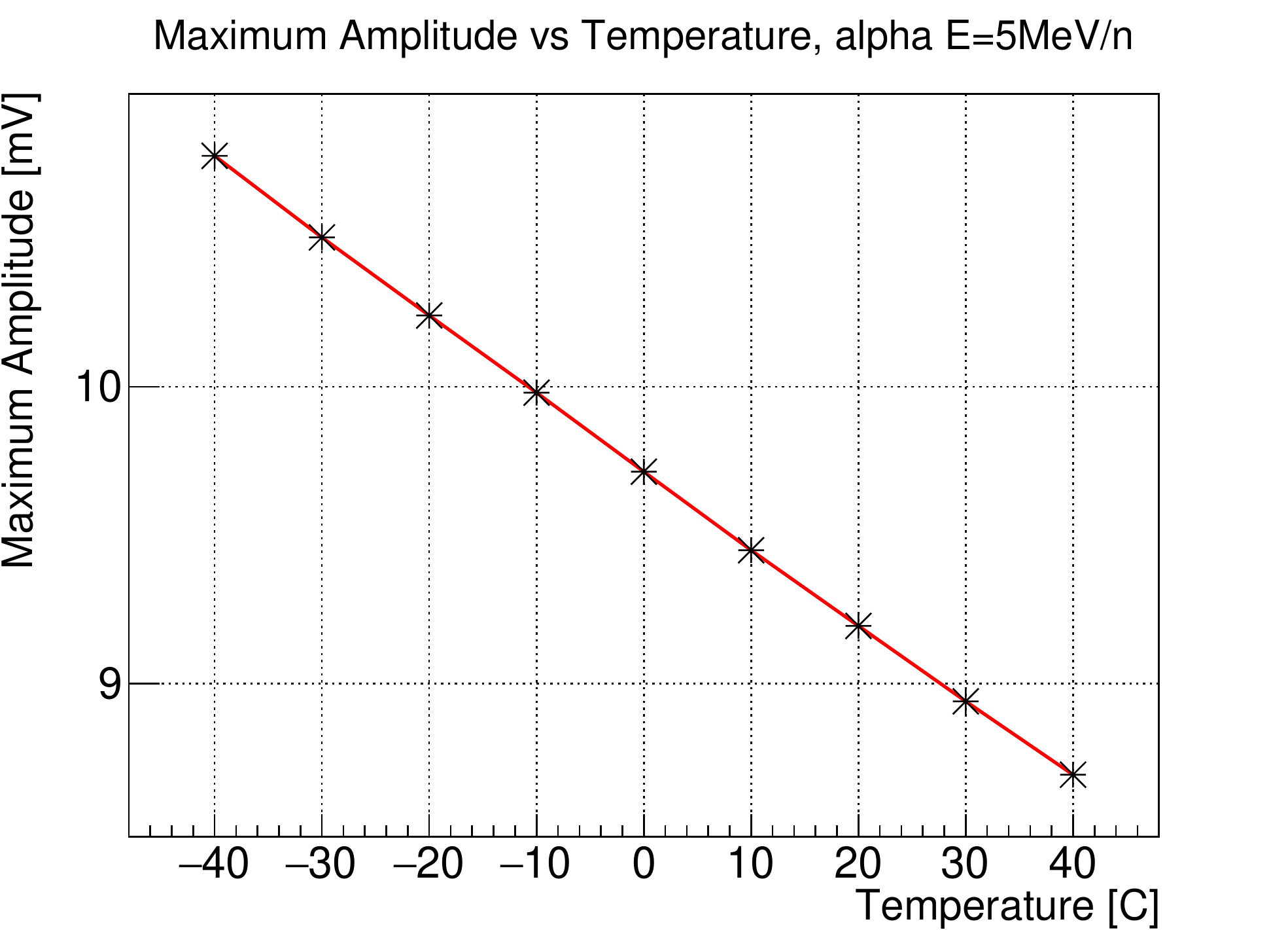} 
        \caption{Amplitude vs temperature}
        \label{fig:amp_temp_stop}
        \end{subfigure}
    \begin{subfigure}{.5\textwidth}
        \centering
        \includegraphics[width=1\textwidth]{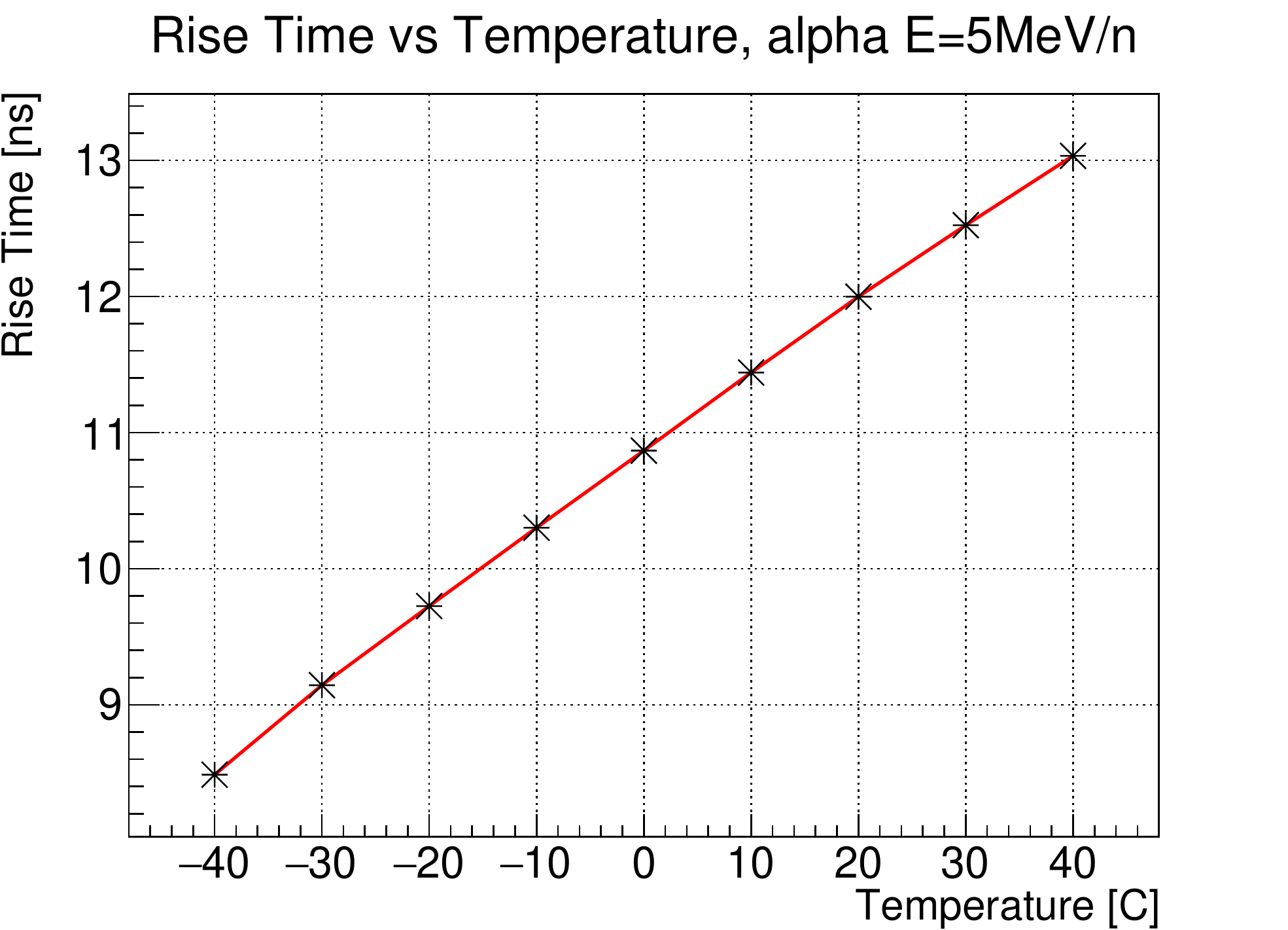} 
        \caption{Rise time vs temperature}
        \label{fig:time_temp_stop}
    \end{subfigure}
    \begin{subfigure}{\textwidth}
        \centering
        \includegraphics[width=0.5\textwidth]{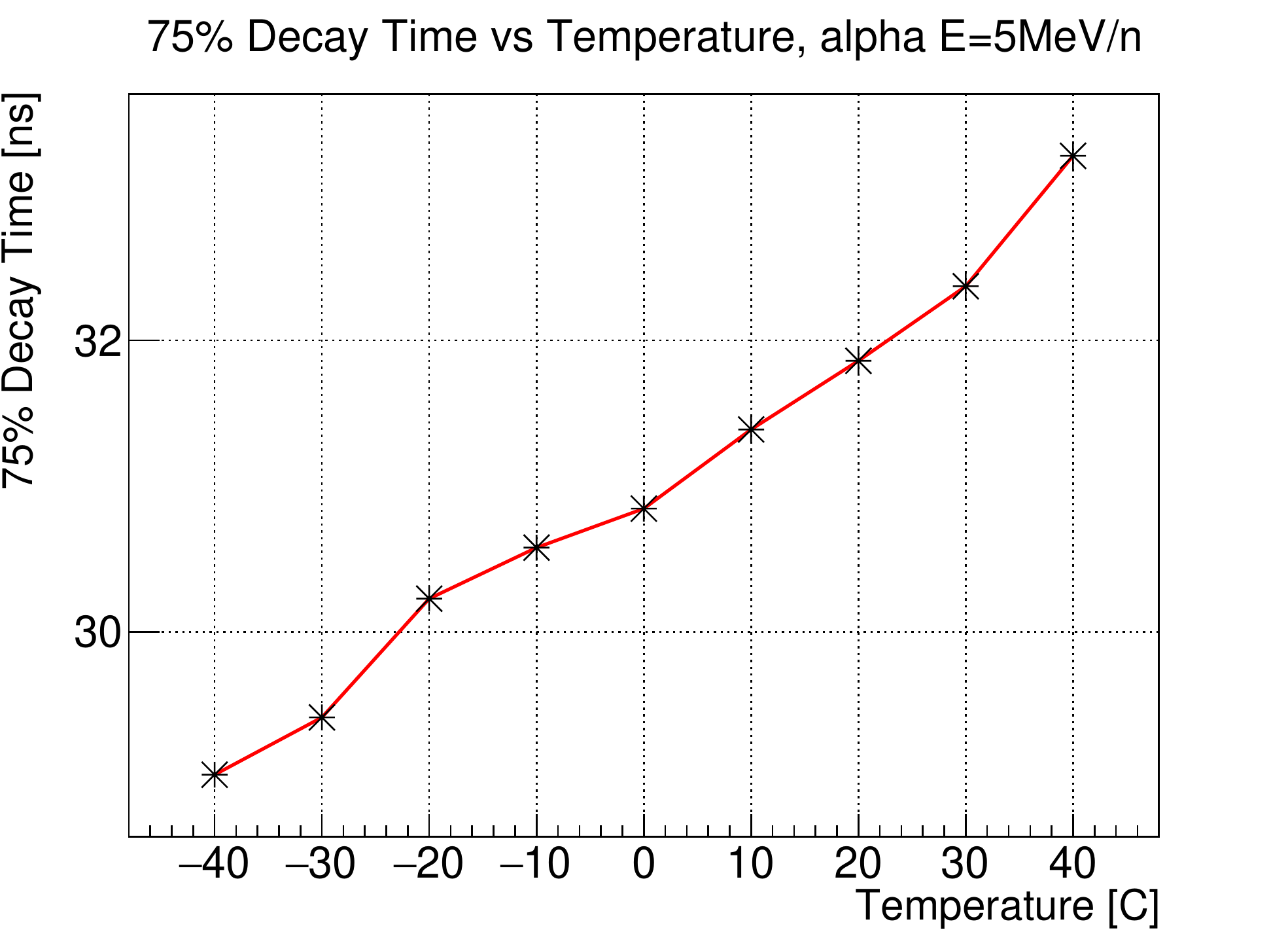} 
        \caption{75\% Decay time vs temperature}
        \label{fig:temperature_75Dec}
    \end{subfigure}
    \caption{Effect of the temperature on the signal key characteristics}
    \label{fig:characteristics_vs_temperature}
\end{figure}

 Fig. \ref{fig:characteristics_vs_temperature} shows the dependence of the main signal characteristics such as amplitude (\ref{fig:amp_temp_stop}), rise time (\ref{fig:time_temp_stop}), and 75\% decay time (\ref{fig:temperature_75Dec}) on temperature. As can be seen, the key signal characteristics are proportional to temperature and thus can be corrected on-board if the ambient temperature is known.\par
   
In conclusion, the main source of uncertainty on signal characteristics is the particle incident angle.
 While electronics noise affects light ions such as hydrogen and helium more, the particle incident angle contributes more uncertainty for the heavier ions such as iron. It is foreseen for the current AGILE design, to limit the angular acceptance to $40^{\circ}$ field of view.  In addition, the temperature influence can be corrected according to on-board measurements. The AGILE instrument will have sensors at key points to measure the local ambient temperature.

\section{Future Improvements}

There are multiple ways to improve the performance of the PSD method, in particular: 
\begin{itemize}
    \item A detailed analysis of not only the signals from the layers when particles stop, but also from the layers where particles are passing through. This can extend the energy range of the identified particles.
    \item Using different variations of the detector medium (Si) doping and its orientation (such as letting particles enter the detector from its n-side) can possibly reduce the width of the ``dead zone" of the detector for which the signals from all the particles stopping there will have the same rise time.
    \item Using the detector layers with different thickness and/or adding absorbers can extend the energy range of the identified particles and make the discrimination method more sensitive for specific ranges of energy of interest for each particle.
    \item Adding a discrimination method for the incident angle of incoming particles, such as use of segmented detectors (e.g., bulls-eye, guard ring etc). Another approach would be to use shaped detectors (e.g. the front detectors of PET instrument onboard SAMPEX) \citep{cook1993pet}.
\end{itemize}

\section{Conclusion}
For space-based instruments, the identification of ions and their energy measurements can be accomplished by means of using real-time on-board pulse shape discrimination techniques. AGILE proposes a novel method to precisely identify charged particles using multiple layers of silicon detectors along with custom made read-out electronics and fast modern samplers. It relies on the identification of the particle charge and mass using the rise time and maximum amplitude of the signal produced in the stopping silicon layer. Subsequent energy estimation is based on the maximum amplitude of this same signal.
The main advantage of this new method with respect to the widely used $\Delta E-E$ technique is in use of both amplitude and timing information of the signal and the possibility of particle discrimination using only the detector layer where the particle stops.

Based on the results of detailed simulations, the PSD method is expected to provide a robust discrimination of a large variety of ions (from H to Fe) and their energy determination with a resolution of less than $\sim 5\%$ and an ID discrimination efficiency up to $100\%$ within the range of its energy acceptance.
\par
Since induced signal amplitude  significantly differs between light and heavy ions (few orders of magnitude), AGILE read-out electronics will be adapted depending upon measurement emphasis on light vs. heavy ions by having different chains of amplification.  \par

Using this method with three layers of silicon, we can measure, as an example, oxygen ions up to $\sim 22$ MeV/n and discriminate them from other ions such as iron which we can measure up to $\sim 45$ MeV/n.

In order to verify and to consolidate the simulation conclusions and to calibrate the instrument, AGILE collaboration plans to schedule a test beam campaign at Brookhaven National Laboratory, using a large variety of ions in a wide energy range.

\section{Acknowledgements}
This work is supported by Heliophysics Technology and Instrument Development for Science (HTIDS) ITD and LNAPP program part of NASA Research Announcement (NRA) NNH18ZDA001N-HTIDS for Research Opportunities in Space and Earth Science – 2018 (ROSES-2018).
\bibliographystyle{elsarticle-num-names}
\bibliography{main.bib}

\appendix

\section{Detector Response Simulation}
\label{section:simulation}
\begin{figure}[H]
    \centering
    \includegraphics[width=1\textwidth]{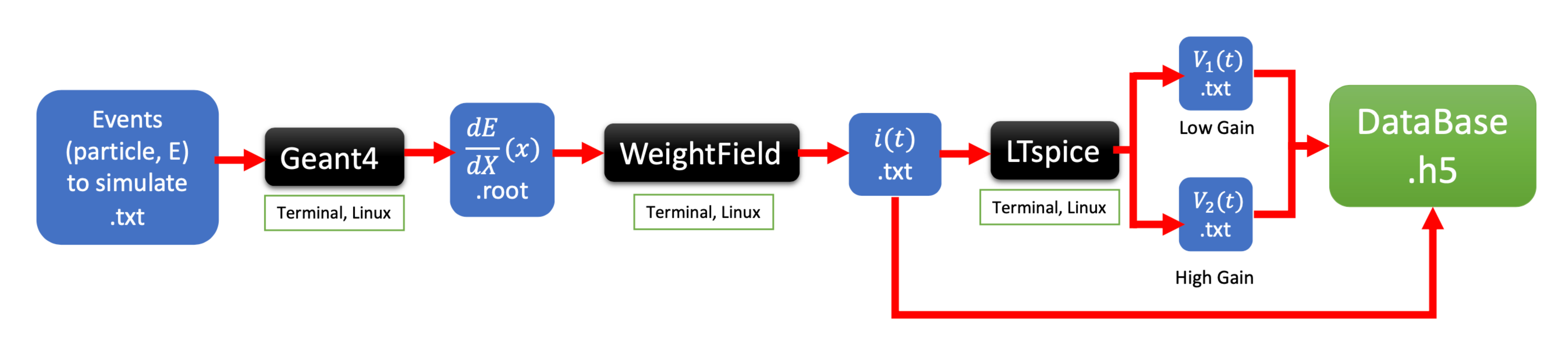}
    \caption{Full chain of simulation}
    \label{fig:full_chain}
\end{figure}

The expected performance of the method described in these studies was obtained by means of detailed simulations consisting of three main stages (Fig. \ref{fig:full_chain}). 



Various ions (H-Fe) in the energy range of (1-100)MeV/n were used as input events. The first stage in the chain is the Geant4 toolkit which simulates the energy deposition profile in all layers of the detector. Then this information is passed to Weightfield2 software that simulates the detector response, which is then processed by LTspice to obtain the read-out electronics output. \par 
The three following subsections focus on each of these steps. In order to validate the simulation results a beam test with a large variety of ions in a wide energy range is planned.

\subsection{Energy Loss (GEANT4 \citep{geant4})}

The geometry of the instrument was simplified to only the main components that affect the shape of the output signals. This includes not only the layers of sensitive silicon detectors but also the aluminium metalization of the detectors. In addition, a primary kapton absorber is placed in front of the stack of detectors to stop low-energy particles  (E$<$ 1 MeV/n) not of interest for AGILE. \par

One of the three mandatory (and most important) user classes of the Geant4 toolkit is \textit{Physics list} which needs to be chosen according to the application. In these studies a specialized low energy option well adapted for space applications called the Electromagnetic standard \textit{Physics list} (option 4) was used. To obtain high spatial resolution, an energy loss profile with a 1\micro m accuracy was chosen. In this option, the simulation of energy loss in silicon mostly relies on the Bethe-Bloch formula \citep{van2006Ident}, which is complemented by the shell correction (Bischel Model \citep{bichsel1988straggling}) to get more physically accurate results.

\subsection{Output Current (Weightfield2) \cite{WF2}}
To simulate the detector response, the Weightfield2 software was used. Originally this tool was developed to work with minimum ionizing particles (MIPs) but it was modified to operate with a wide variety of particles and energy ranges. Various configurations of the silicon detector can be chosen for pulse shape discrimination, in this study a 300 \micro m p-type bulk detector with a back side n-type electrode was chosen as the default. The depletion voltage in this case is about 60V, the bias voltage used is 110V. In order to account for the total fluctuation of the number of carriers in the silicone detector, a Fano factor \cite{fano1947ionization} was taken to be equal to 0.1.

\subsection{Read-Out Electronics}
\begin{figure}[H]
    \centering
    \includegraphics[width=0.8\textwidth]{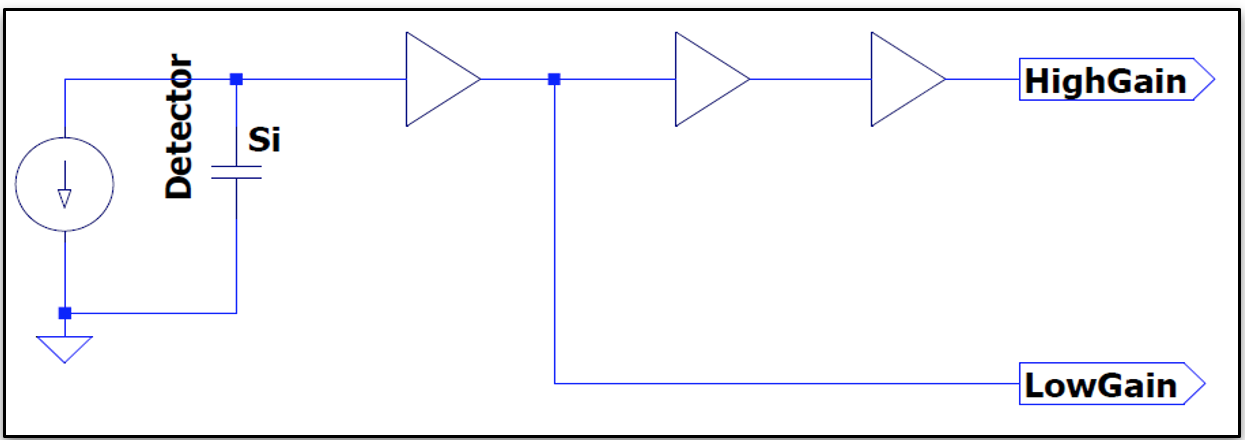}
    \caption{Two Gains Amplifier}
    \label{fig:2gain_amp}
\end{figure}

Since the AGILE instrument registers a large variety of particles in a wide energy range the amplitude of the detector output signals varies significantly: $10^{-7}$A - $10^{-2}$A (almost 5 orders of magnitude). In order to address this while maintaining a high signal-to-noise ratio two stages of amplification (low and high gain) are used (Fig. \ref{fig:2gain_amp}).

 In the final stage of the chain, LTspice software \citep{ltspice} is used to simulate the output signals from both low and high gains of the amplifier (V(t)), which are then stored in a specific database encoded in HDF5 format along with the detector response (I(t)).

\section{AGILE Scientific Goals}
\label{agile}
The scientific goals and instrument implementation of AGILE are described in detail in a companion paper (Kanekal et al., JGR to be submitted).  In brief, these include (but are not limited to):
\begin{itemize}
    \item studies of the energization and transport of solar energetic particles (SEPs) for better understanding of two types of SEP events: gradual (proton-reach) and impulsive (\ce{^{3}He-reach}) \citep{reames99};
    \item examination of trapped and transient populations of anomalous cosmic rays which can be used in studies of: dynamics of energetic particles within the solar system, general properties of the heliosphere, nature of interstellar material \citep{cummings93};
    \item unambiguous estimation of the relativistic electrons in the Inner Van Allen Belt in the presence of a large proton population \citep{li2015};
    \item understanding of space weather events with regards to harmful effects of particle radiation to humans and human assets in space \citep{mewaldt2005};

\end{itemize}

\subsection{Particles' Fluxes Estimation}

In order to adjust the hardware and the PSD algorithms, it is crucial to have information about the fluxes of different particles in the regions where AGILE will be operating. The most important "contributors" in these fluxes are Galactic Cosmic Rays (GCRs), Anomalous Cosmic Rays (ACRs), and Solar Energetic Particles (SEPs)

{\bf GCRs + ACRs} \par
\begin{figure}[!htp]
    
        \centering
        \includegraphics[width=1\textwidth]{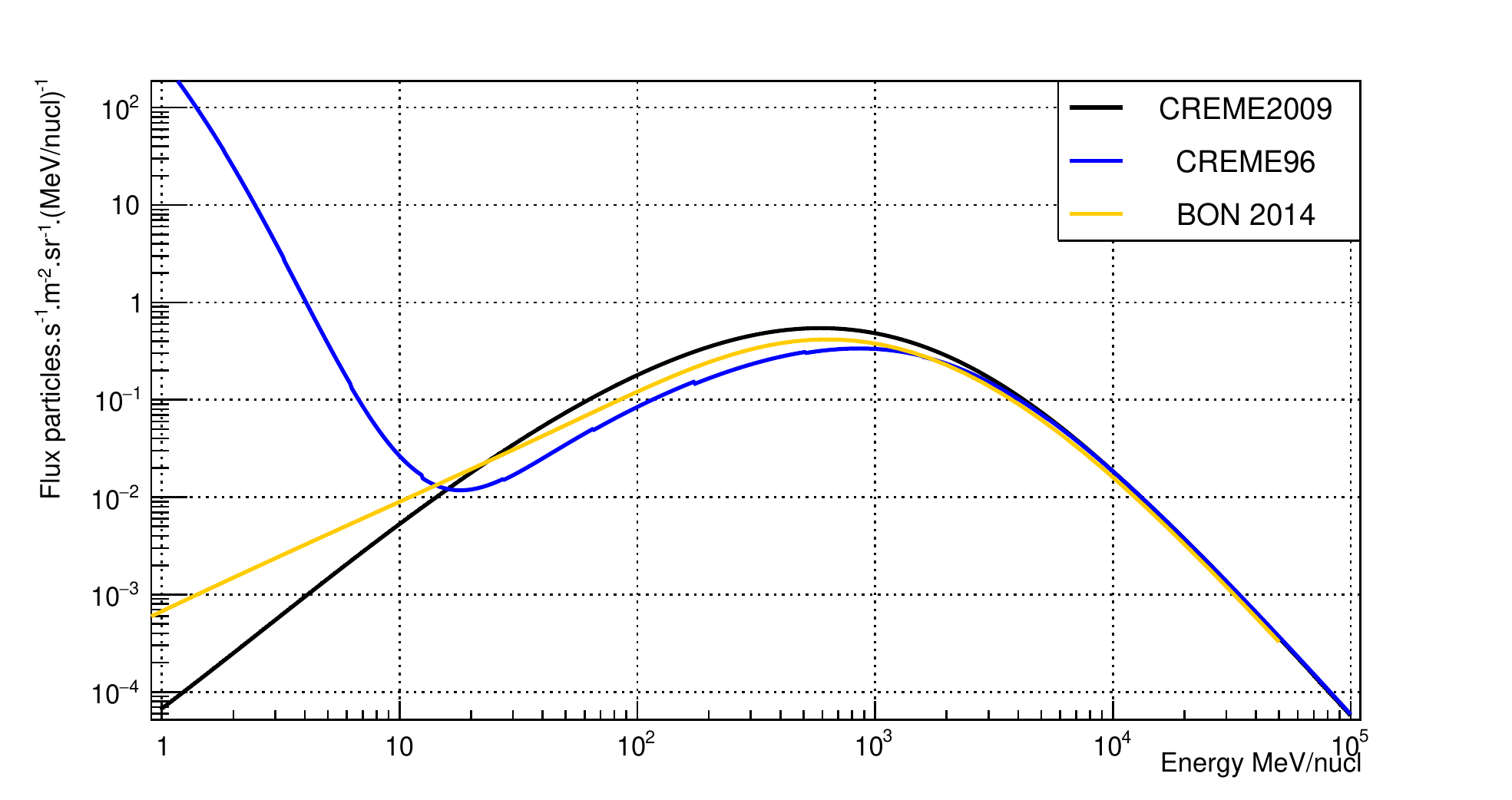} 
        \caption{Hydrogen GCR models (solar maximum, interplanetary space) }
        \label{fig:gcrH}
\end{figure}
\begin{figure}[!htp]
        \centering
        \includegraphics[width=1\textwidth]{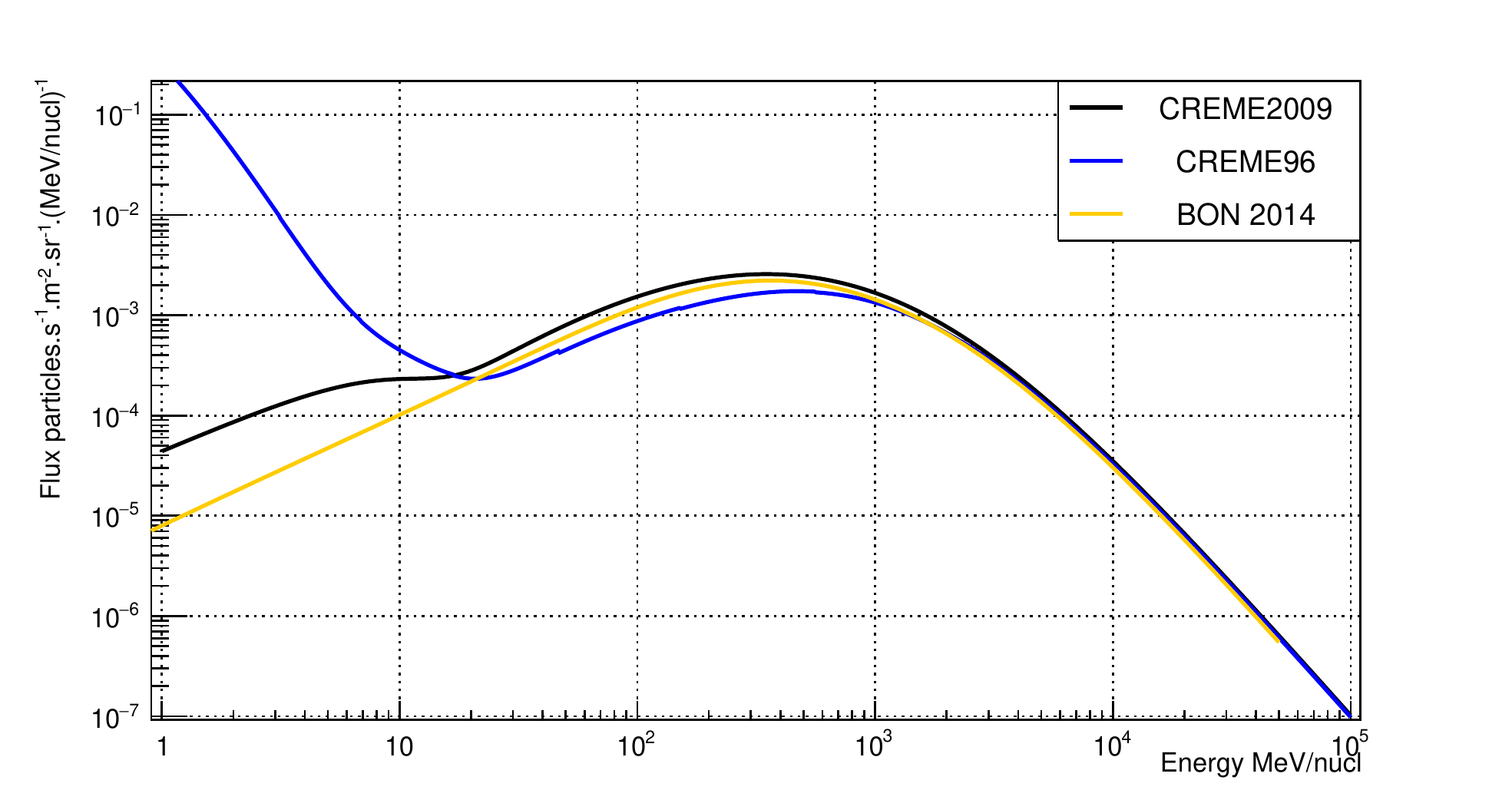} 
        \caption{Oxygen GCR models (solar maximum, interplanetary space)}
        \label{fig:gcrO}
 
\end{figure}

To estimate the flux of GCRs a few different models have been used: 
\begin{itemize}
    \item BON \cite{oneill2010};
    \item CREME96 and its update CREME2009 \cite{tylka97};
    \item ISO15390.
\end{itemize}
These models can estimate the flux for different ions with Z=1 to 28, both in the interplanetary space as well as inside the Earth atmosphere. As an example Fig. \ref{fig:gcrH} and Fig. \ref{fig:gcrO} show the GCRs fluxes for two ions (H and O) obtained using various models. The models can vary with the solar activity, e.g. solar minimum corresponds to a cosmic rays maximum, and solar maximum corresponds to a cosmic rays minimum.

\begin{figure}[!htp]
    \centering
    \includegraphics[width=0.95\textwidth]{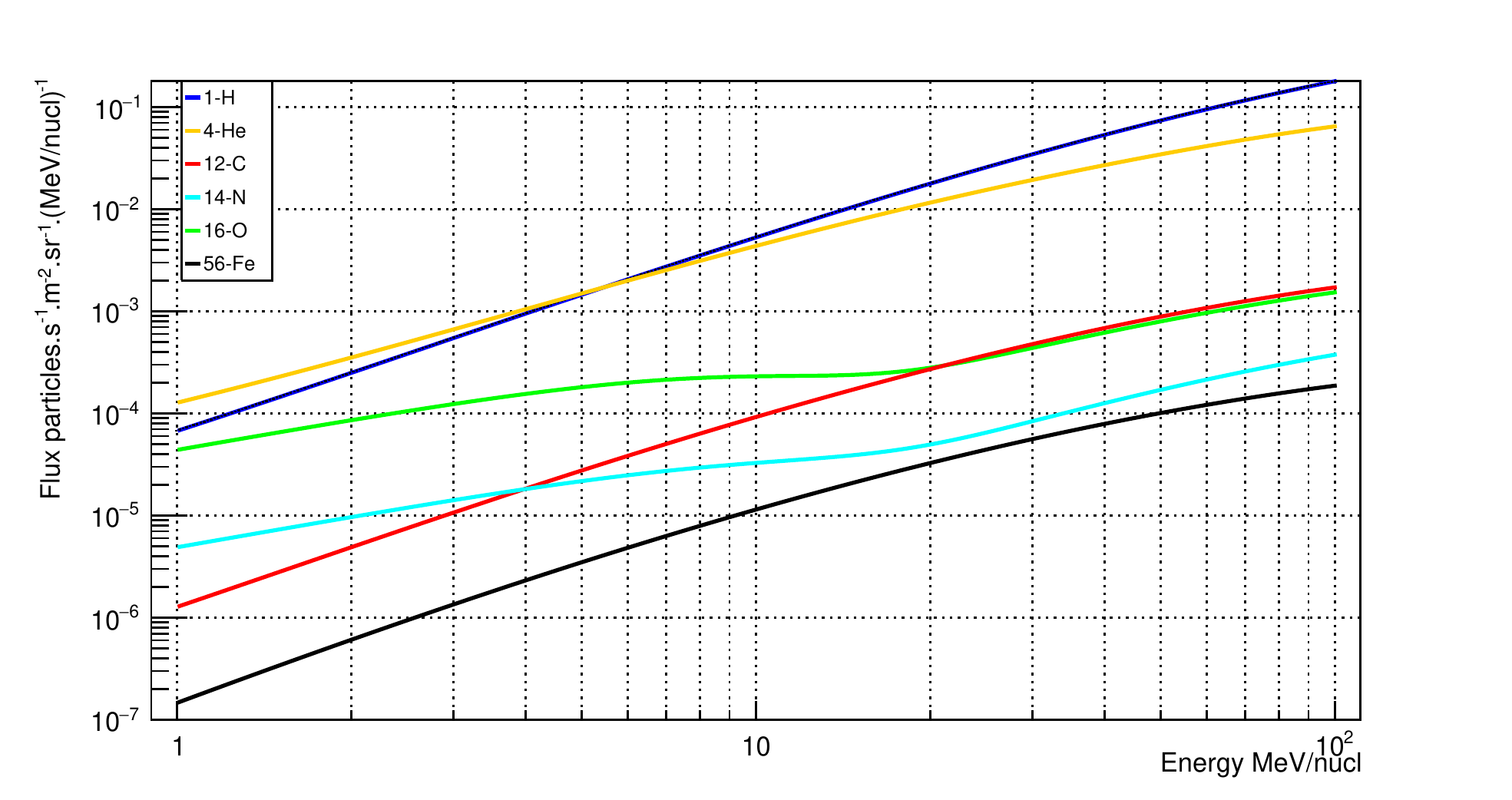}
    \caption{CREME2009 model estimation of GCR+ACR in the interplanetary space (1AU) during a solar maximum, for Z=1 to 26. The increase at about 10 Mev/nuc for O and N among others correspond to Anomalous Cosmic Rays}
    \label{fig:cremeacr}
\end{figure}

In addition to GCRs, CREME models also take into account the ACRs. Fig. \ref{fig:cremeacr} shows the estimated fluxes of GCRs+ACRs for various ions. \par
{\bf SEPs} \par
When specific event measurements do no exist, characterizing the peak flux for a single SEP event is a challenging task since some (rare) events may have much higher flux compared to others. To address this we introduce the probability $p_E(f)$ of having a flux over $f$ for each particle and energy. The SAPPHIRE \cite{jiggens2018} model, and Spenvis database \cite{spenvis} give this distribution. Then, for a given probability, we fit the energy spectra with the formula: \\

 For E$<(\gamma_b-\gamma_a)$.E$_0$:\\
 \begin{equation} 
 f(E)=K.E^{-\gamma_a}.exp(-E/E_0) \\
 \end{equation}
 For E$>(\gamma_b-\gamma_a)$.E$_0$:\\
 \begin{equation}
 f(E)=K.E^{-\gamma_b}.((\gamma_b-\gamma_a).E_0)^{\gamma_b-\gamma_a}.exp(\gamma_a-\gamma_b) \\
 \end{equation}

\begin{figure}[!htp]
    \centering
    \includegraphics[width=0.9\textwidth]{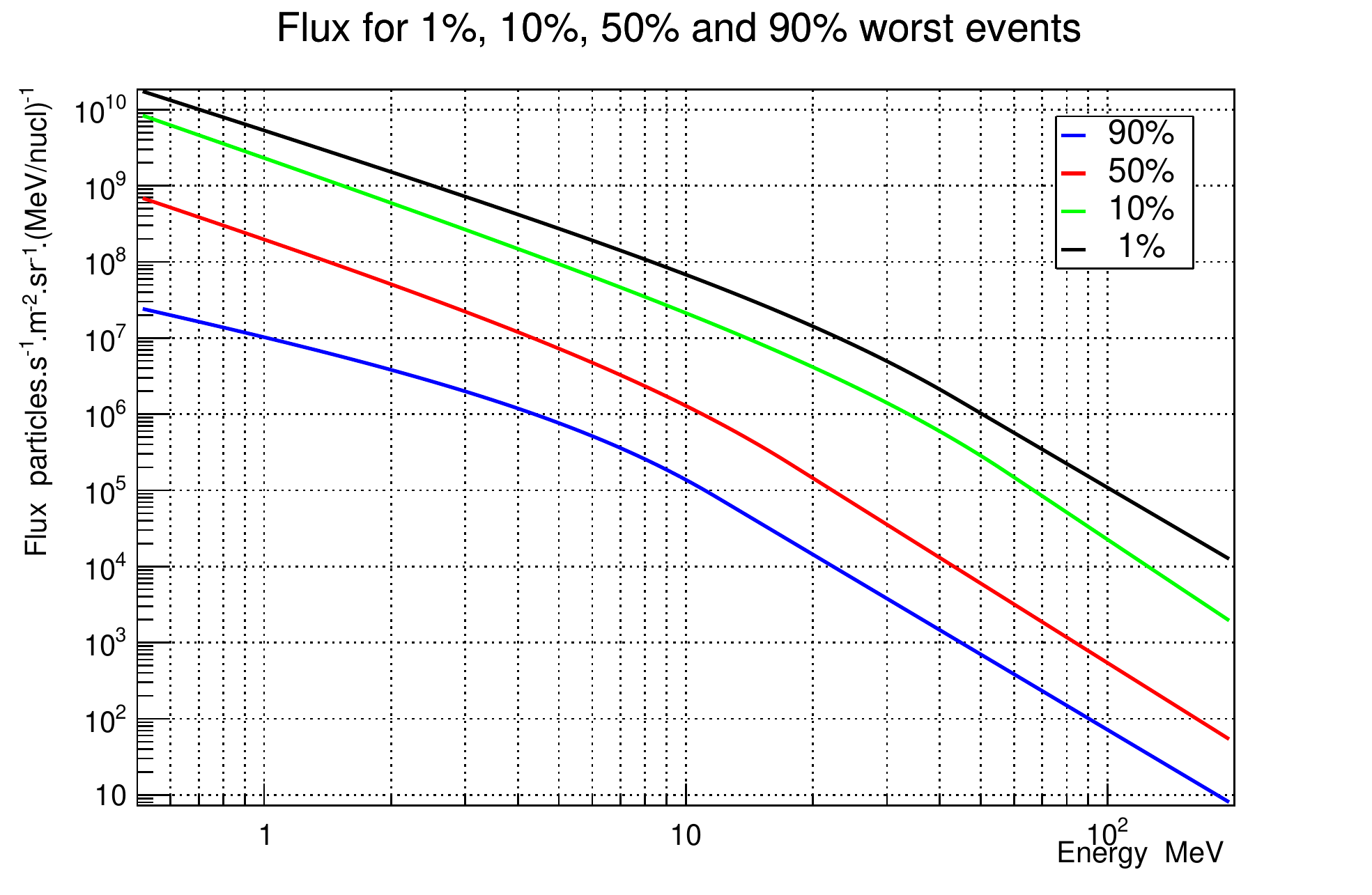}
    \caption{SEP proton flux (circular orbit, altitude 1000 km, inclination 90\degree) with probability of having higher flux of 1\% (black), 10\% (green), 50\% (red), and 90 \% (blue).}
    \label{fig:hsep}
\end{figure}

\begin{figure}[!htp]
    \centering
    \includegraphics[width=0.9\textwidth]{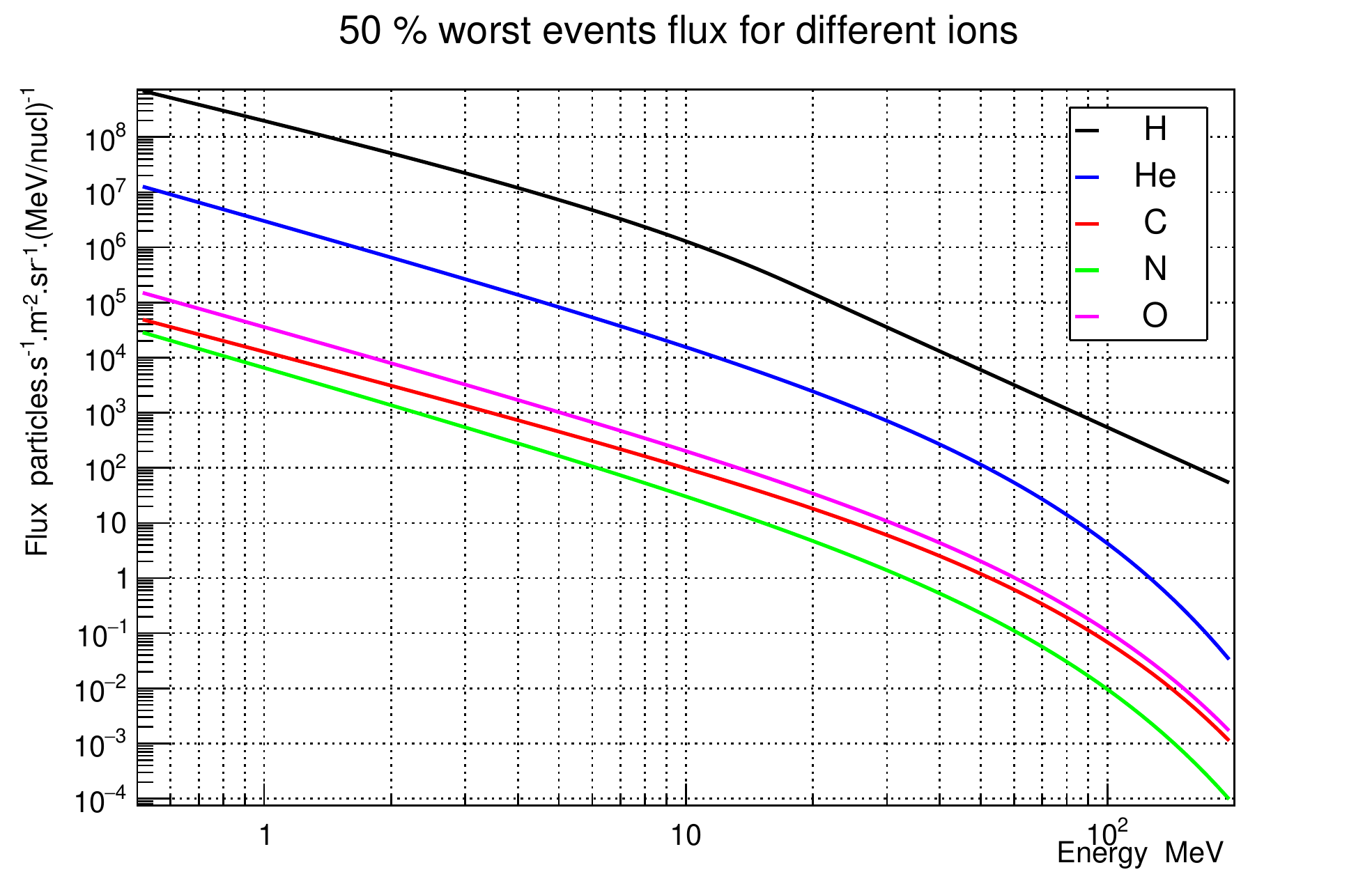}
    \caption{SEP flux (circular orbit, altitude 1000 km, inclination 90\degree) with probability of having higher flux of 50\% for a few species.}
    \label{fig:sep50}
\end{figure}

All these values will be used as parameters in AGILE in order to estimate the data rate production during normal operation or in events of important fluxes.












\end{document}